\documentclass[sn-nature,iicol,pdflatex,a4paper]{sn-jnl-arxiv}

\usepackage{amsmath,amssymb,amsfonts}
\usepackage{amsthm}
\usepackage{mathrsfs}
\usepackage[title]{appendix}
\usepackage{textcomp}
\usepackage{manyfoot}
\usepackage{booktabs}

\usepackage[utf8]{inputenc}
\usepackage{microtype}
\usepackage{xcolor}
\usepackage{xspace}
\usepackage{amssymb}
\usepackage[frozencache=true]{minted}
\usepackage{makecell}
\usepackage{graphicx}
\usepackage{hyperref}
\usepackage{pifont}
\usepackage{subcaption}
\usepackage{multicol}
\usepackage{relsize}
\usepackage{tabularx}

\renewcommand{\listingscaption}{\bf Listing}

\DeclareUnicodeCharacter{2713}{\checkmark}

\title{Potential of the Julia programming language for high energy physics computing}
\subtitle{{\em Published in }Computing and Software for Big Science {\em as \href{https://doi.org/10.1007/s41781-023-00104-x}{doi:10.1007/s41781-023-00104-x}}}

\author[1]{Jonas Eschle}
\author[2]{Tam\'as G\'al}
\author[3]{Mos\`e Giordano}
\author*[4]{Philippe Gras}\email{philippe.gras@cern.ch}
\author[5]{Benedikt Hegner}
\author[6]{Lukas Heinrich}
\author[7,8]{Uwe Hernandez Acosta}
\author[6]{Stefan Kluth}
\author[9]{Jerry Ling}
\author[5]{Pere Mato}
\author[10,11]{Mikhail Mikhasenko}
\author[12]{Alexander Moreno Briceño}
\author[13]{Jim Pivarski}
\author[5]{Konstantinos Samaras-Tsakiris}
\author[6]{Oliver Schulz}
\author[5]{Graeme Andrew Stewart}
\author[14,15]{Jan Strube}
\author[13]{Vassil Vassilev}

\newcommand{\affilfontsize}{\small}

\affil[1]{\affilfontsize\orgname{University of Zurich}, \orgaddress{\city{Zürich}, \country{Switzerland}}}
\affil[2]{\affilfontsize\orgname{Erlangen Centre for Astroparticle Physics, Friedrich-Alexander-Universität}, \orgaddress{\city{Erlangen-Nürnberg}, \country{Germany}}}
\affil[3]{\affilfontsize\orgname{University College London}, \orgaddress{\city{London}, \country{United Kingdom}}}
\affil*[4]{\affilfontsize\orgname{IRFU, CEA, Universit\'e Paris-Saclay}, \orgaddress{\city{Gif-sur-Yvette}, \country{France}}}
\affil[5]{\affilfontsize\orgname{CERN, European Organization for Nuclear Research}, \orgaddress{\city{Geneva}, \country{Switzerland}}}
\affil[6]{\affilfontsize\orgname{Max-Planck-Institut für Physik}, \orgaddress{\city{Munich}, \country{Germany}}}
\affil[7]{\affilfontsize\orgname{Center for Advanced Systems Understanding}, \orgaddress{\city{G\"orlitz}, \country{Germany}}}
\affil[8]{\affilfontsize\orgname{Helmholtz-Zentrum Dresden-Rossendorf}, \orgaddress{\city{Dresden}, \country{Germany}}}
\affil[9]{\affilfontsize\orgname{Laboratory for Particle Physics and Cosmology, Harvard University}, \orgaddress{\city{Cambridge}, \state{MA}, \country{USA}}}
\affil[10]{\affilfontsize\orgname{ORIGINS Excellence Cluster}, \orgaddress{\city{Garching}, \country{Germany}}}
\affil[11]{\affilfontsize\orgname{Ludwig-Maximilians-Universit{\"a}t}, \orgaddress{\city{Munich}, \country{Germany}}}
\affil[12]{\affilfontsize\orgname{Universidad Antonio Nari\~no}, \orgaddress{\city{Ibagué}, \country{Colombia}}}
\affil[13]{\affilfontsize\orgname{Princeton University}, \orgaddress{\city{Princeton}, \state{NJ}, \country{USA}}}
\affil[14]{\affilfontsize\orgname{Pacific Northwest National Laboratory}, \orgaddress{\city{Richland}, \state{WA}, \country{USA}}}
\affil[15]{\affilfontsize\orgname{University of Oregon}, \orgaddress{\city{Eugene}, \state{OR}, \country{USA}}}

\newcommand\codename[1]{\texttt{#1}}
\newcommand\juliapkg[1]{\codename{#1}}

\newcommand{\Rplus}{\protect\hspace{-.1em}\protect\raisebox{.35ex}{\smaller{\smaller\textbf{+}}}}
\newcommand{\Cpp}{\mbox{C\Rplus\Rplus}\xspace}
\newcommand\fname[1]{\texttt{#1}}

\abstract{Research in high energy physics (HEP) requires huge amounts of computing and storage, putting strong constraints on the code speed and resource usage. To meet these requirements, a compiled high-performance language is typically used; while for physicists, who focus on the application when developing the code, better research productivity pleads for a high-level programming language. A popular approach consists of combining Python, used for the high-level interface, and C++, used for the computing intensive part of the code. A more convenient and efficient approach would be to use a language that provides both high-level programming and high-performance. The Julia programming language, developed at MIT especially to allow the use of a single language in research activities, has followed this path. In this paper the applicability of using the Julia language for HEP research is explored, covering the different aspects that are important for HEP code development: runtime performance, handling of large projects, interface with legacy code, distributed computing, training, and ease of programming. The study shows that the HEP community would benefit from a large scale adoption of this programming language. The HEP-specific foundation libraries that would need to be consolidated are identified.}

\unnumbered

\begin{document}

\maketitle

\section{Introduction}

High throughput computing plays a major role in high energy physics (HEP) research. The field requires the development of sophisticated computing codes, which are continuously evolving in the course of the research work. Computing grids, connecting computer centers all around the world, are required to process the experiments' data~\cite{bib:WLCG}. Computer algebra systems and high performance computers are used to build new models and to calculate particle production cross sections.

\begin{figure}[b]
  \centering
  \includegraphics[width=\linewidth]{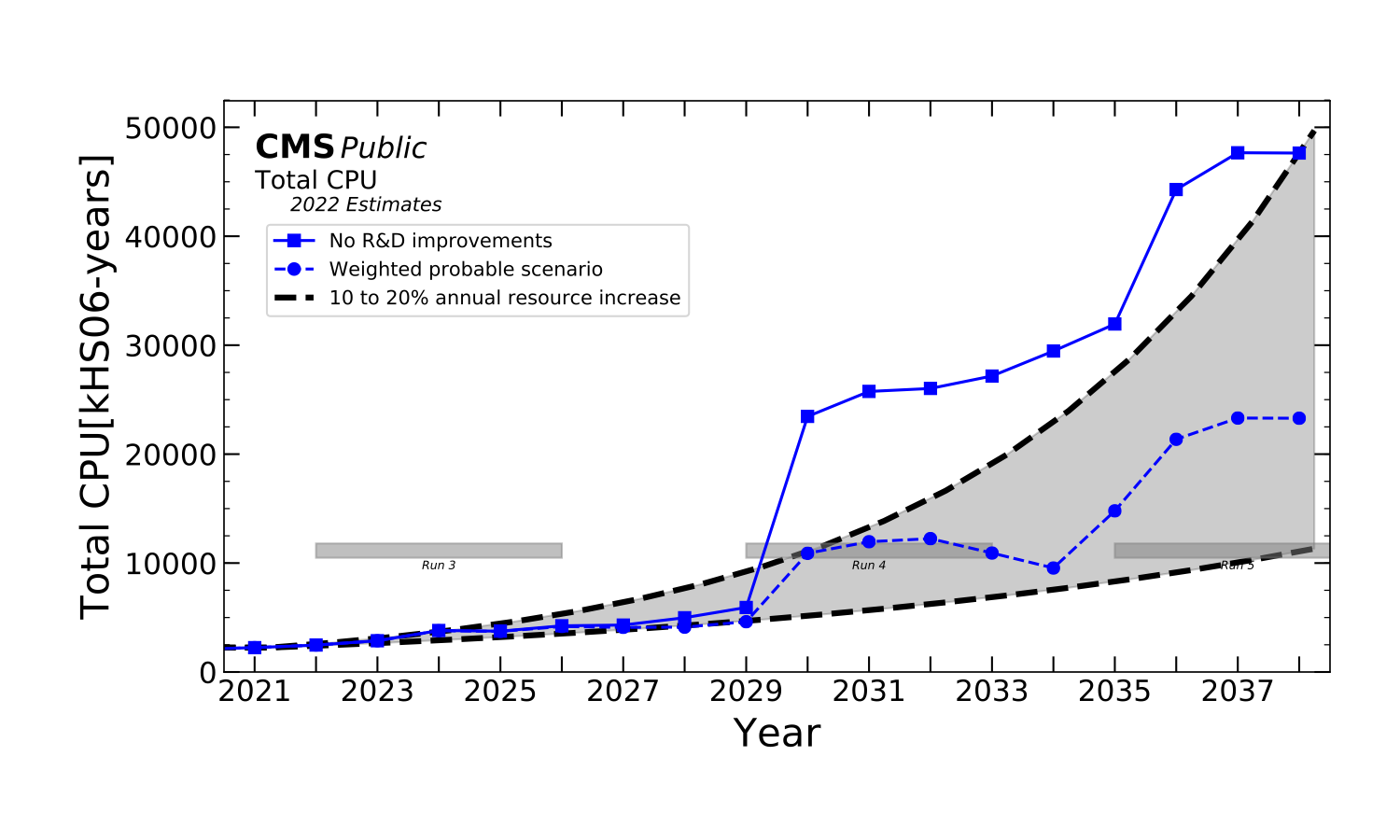}\\
  \includegraphics[width=\linewidth]{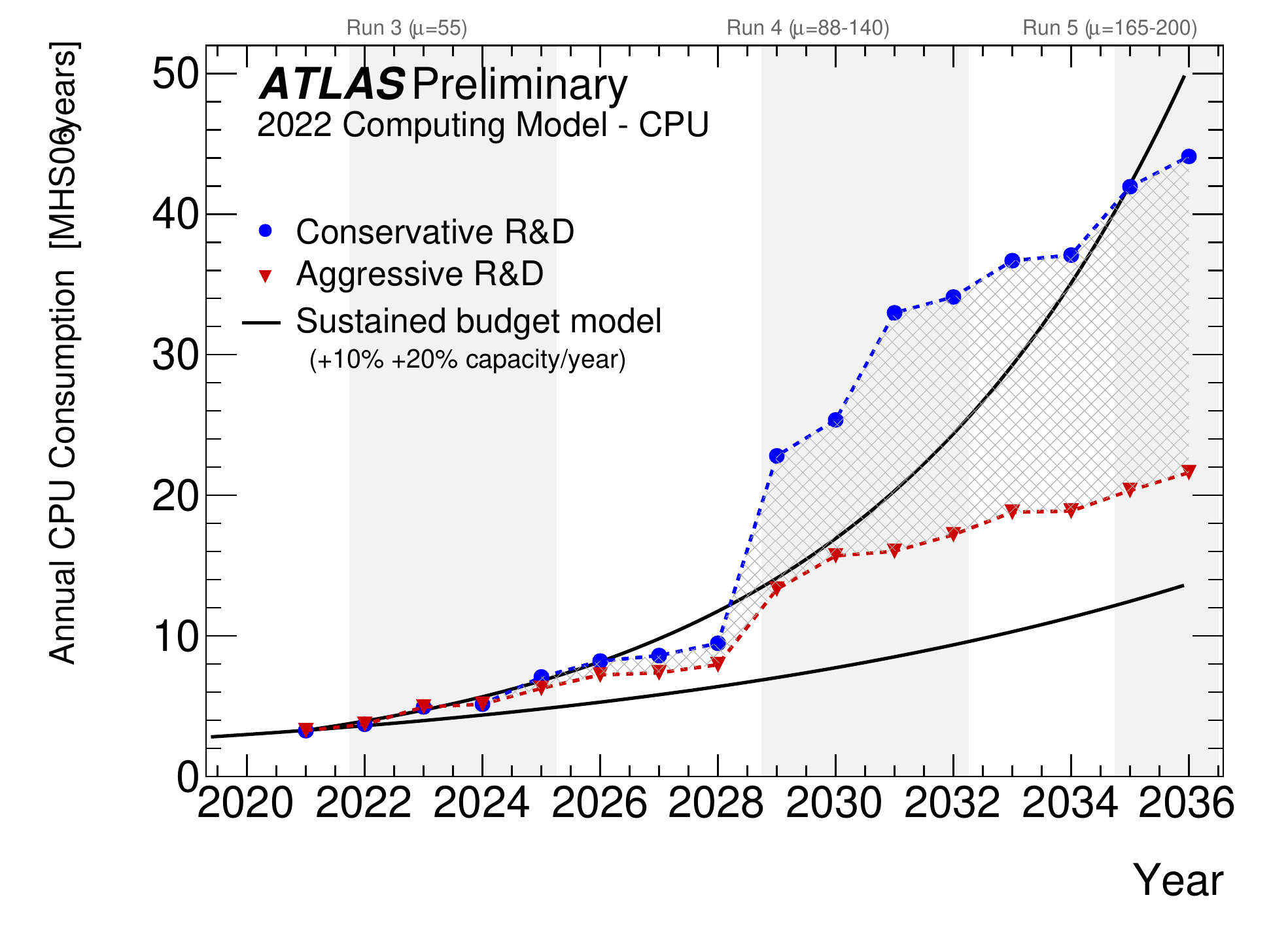}
  \caption{Estimated CPU required by the CMS (top) and ATLAS (bottom) experiments for LHC and HL-LHC~\cite{HEPSoftwareFoundation:2017ggl,Sexton_Kennedy_2018}.}
  \label{HL-LHC-computing}	
\end{figure}

Fig.~\ref{HL-LHC-computing} shows the expected needs for the ATLAS and CMS experiments~\cite{Collaboration:2802918,Software:2815292} at the Large Hadron Collider (LHC)~\cite{Evans:2008zzb} and its successor, the high-luminosity LHC (HL-LHC)~\cite{Apollinari:2017lan}, together with the estimated planned resources. A data processing improvement from R\&D is required for HL-LHC to fit within the planned resources, which total $20\cdot 10^9$ HS06\footnote{https://www.spec.org/cpu2006/} units of CPU resource.  


The need to reconcile high performance with fast development has led to the development of a \Cpp{} interpreter~\cite{cling} that provides the convenience of a read-eval-print-loop (REPL) interactive experience, also known as programming shell, that supports just-in-time compilation, and allows the use of the same programming language for compiled and interpreted code. The same analysis framework \codename{ROOT}~\cite{Brun:1997pa,ANTCHEVA20092499} can then be used with compiled code and interactively. In addition to the REPL, ROOT supports Jupyter notebooks, which are another convenient method for interactive use. The shortcoming of this approach is that the use of a complex programming language is not optimal for easy and fast coding. For this reason, another approach that consists of using two languages, one optimal for fast development, typically Python, and one optimal for high performance, typically \Cpp, is often adopted.

Using two languages is not ideal: it expands the required area of expertise; it forces the reimplementation, in the high-performance language, of pieces of code originally written with the fast-development language when they do not meet the required performance; and it reduces the reusability of code.

\begin{figure}
	\centering
	\includegraphics[width=\linewidth]{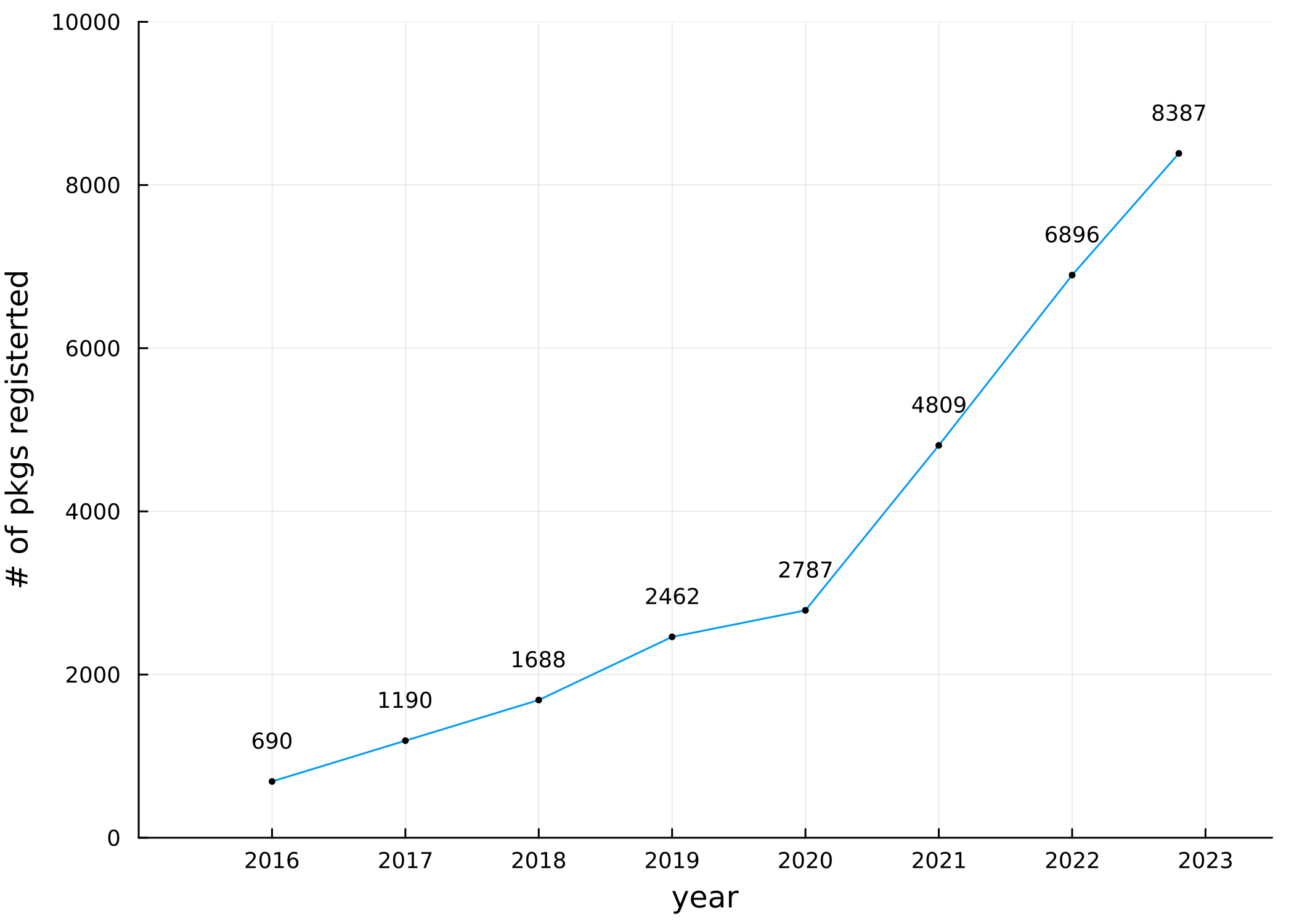}\\
	\includegraphics[width=\linewidth]{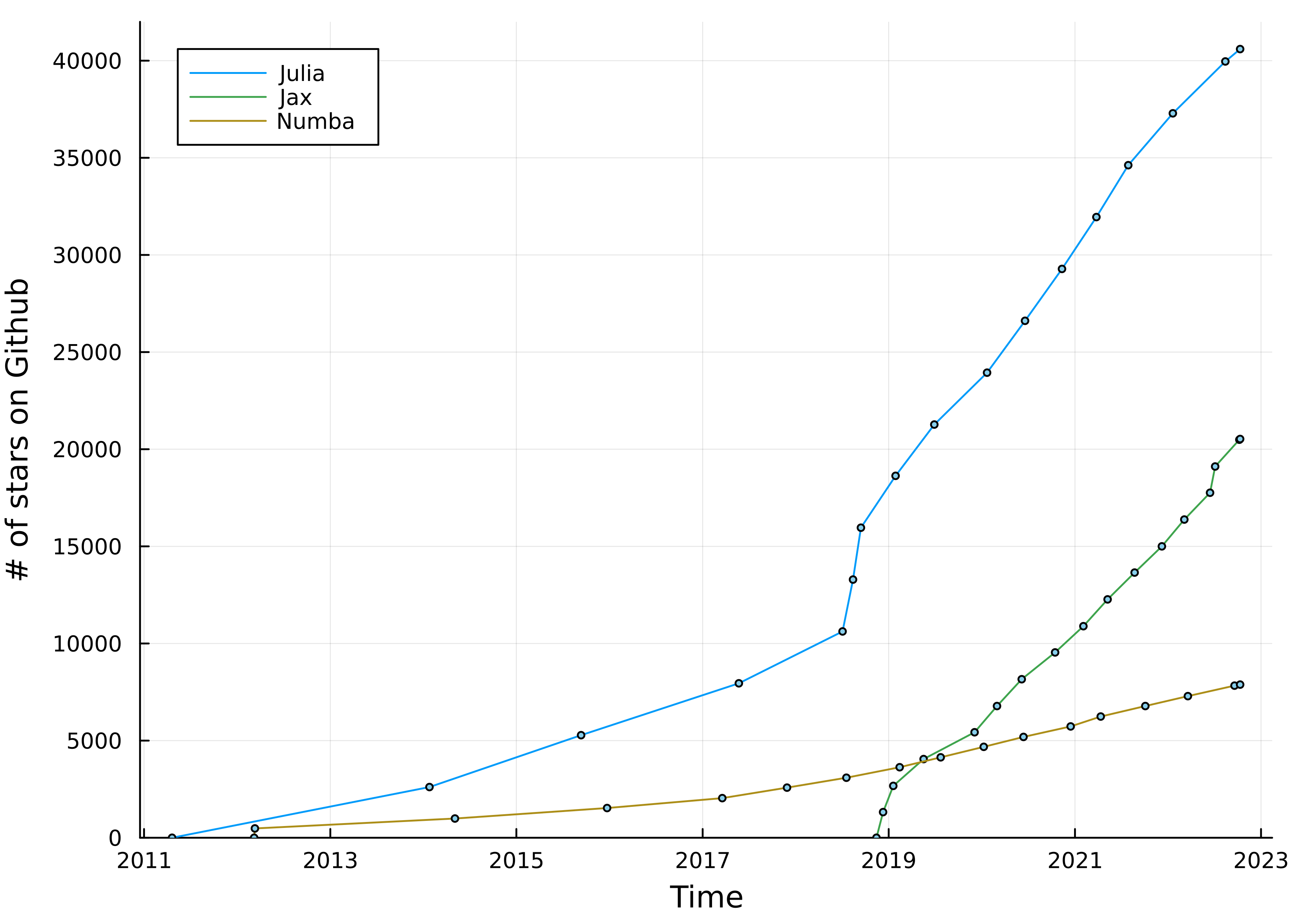}
	\caption{Number of packages registered in the Julia general repository that can be installed by the integrated package manager (top) and of Julia language code GitHub stars (bottom) as function of time. The trend of the star counts is compared with \codename{Numba} and \codename{Jax}.}\label{fig:julia-downloads}
\end{figure}

In 2009, J. Bezanson, A. Edelman, S. Karpinski, and V.~B. Shah imagined a new programming language to address this ``two language problem'' by providing high performance and ease of programming~\cite{bib:julia_freshapproach,10.1145/3276490,bib:juliabirth} simultaneously. It has been a successful approach. The new language, Julia, which has evolved year upon year, is now used by many users.  Julia is a dynamic language, similar to Python, yet with a performance similar to C/\Cpp. As of October 11, 2022, 8387 packages were registered in the Julia general registry~\cite{bib:julia_growth}, which are accessible to Julia's integrated package manager. Fig.~\ref{fig:julia-downloads} shows the rapid growth in the number of packages.


As demonstrated by M.~Stanitzki and J.~Strube~\cite{Stanitzki:2020bnx}, the Julia language is a good alternative to the combination of C++ and Python for HEP data analysis and it fulfils its promise to be an easy, high performance language. This report extends that study. It explores the possible benefits of the adoption of Julia as the main programming language for HEP, in place of C++-Python, in a similar way as happened with the switch from Fortran to \Cpp in the late 90's.

\subsection{The programming language community}

More important than a list of technical features, however, are the culture and interests of a programming language's community, because the language and its implementation will evolve to satisfy those interests. For instance, the Haskell community is focused on language theory, and is unlikely to put much effort into optimization for high performance computing, and the Go community is so focused on language simplicity that they have resisted try-catch logic~\cite{go-error-handling-feedback}. The Julia community's interests are well aligned with HEP, and many Julia users are in the sciences. We see this in some design choices inherited from existing technical languages (Fortran, R, MATLAB, Wolfram), like 1-based indexing, column-major arrays, and built-in N-dimensional arrays, and also in the effort placed on interoperability with other languages: {\tt ccall}, \codename{PyCall.jl}, \codename{RCall.jl}, \codename{MathLink.jl}, and \codename{JavaCall.jl}. Julia is supported by NumFOCUS (like many Python data science projects), and many of its most prominent applications are in numerical computing: NASA spacecraft modeling ~\cite{nasa-julia}; climate science~\cite{climate-julia}; and the Celeste project~\cite{REGIER201989}, which achieved 1.54~Petaflops on the Cori~II supercomputer~\cite{julia-petaflop}, a first for a dynamic language.

\subsection{Key features of Julia}

To locate Julia in the space of programming languages, its primary features are characterized below.

\begin{itemize}
\item A single implementation, rather than an abstract language specification with multiple slightly incompatible implementations. The Julia computing platform is primarily implemented in Julia (most parts), C, and \Cpp(LLVM), and it has a built-in REPL.
  
\item Every function, including those entered interactively, is compiled just-in-time (JIT) using LLVM as the back end. Julia has no virtual machine, and the JIT is eager (always compiles before execution), unlike the tracing/hot spot JIT seen in \codename{LuaJIT}~\cite{luajit} or metatracing~\cite{BOLZ2015408} like \codename{PyPy}~\cite{pypy}.
  
\item Partly thanks to dynamic typing, and also being a JIT language, Julia fully supports: type reflection, source code as a built-in data type, which enables Lisp-like (hygienic) macros\footnote{\url{https://en.wikipedia.org/wiki/Hygienic_macro}}.
  
\item Fast N-dimensional arrays that store elements in-place.
  
\item Multiple dispatch (MD): a function-call invokes the most specific method that matches the type of \emph{all} arguments. While many languages support (opt-in) multiple dispatch (C\#, Common Lisp), Julia is the first language that uses MD as the central paradigm\footnote{Not only Julia lacks OOP by design, it also has highest number of methods and degree of specialization among languages that support MD~\cite{bib:julia_freshapproach}.} while focusing on performance. MD allows for a surprising amount of code reuse and composition among packages that do not know about about each other (often a problem with OOP languages), this is further discussed in the ``\nameref{polymorphsim}'' section.
  
\item Apart from the lack of classes, Julia has a fairly standard mix of imperative and functional programming styles. Immutability is encouraged by default, but mutable structs and arrays are allowed and are frequently used.
  
\item Built-in parallel processing support. Any piece of a program can be marked for execution in parallel. Threads are scheduled globally---allowing a multithreaded function to call other multithreaded function---on available resources without oversubscribing, saving the developer from the burden of taking care of the number of threads. Computing distributed on several computers is supported. Julia code can run natively on GPUs.
  
\item Objects are not reference counted, but are garbage collected. The garbage collector is standard mark-and-sweep, generational (like Java), but non-compacting, so pointers to objects are valid as long as the objects remain in scope.
	
\end{itemize}

The manner in which polymorphism is supported is the most noticeable difference with CBOO programming languages, like \Cpp and Python, and it merits a dedicated discussion.

\subsection{Polymorphism in \texorpdfstring{\Cpp}{C++}, Python, and Julia} \label{polymorphsim}

Polymorphism is the ``ability to provide a single interface to entities of different types''~\cite{stroustrup2018tour, stroustrup1994design}: a polymorphic function will accept arguments of different types. Here we compare Julia with \Cpp and Python due to their prevalence in HEP.

We can distinguish two classes of polymorphism~\cite{10.1145/6041.6042, Strachey}: ad-hoc polymorphism where a different implementation is provided for each set of types, and universal polymorphism, where a single generic implementation is provided for all the sets. 

Function overloading is an example of ad-hoc polymorphism, while \Cpp templates are an example of universal polymorphism. Ad-hoc polymorphism can be combined with universal polymorphism using template specialization: several implementations are provided, while each implementation can be generic, either partially or totally. A particular universal polymorphism is based on subtypes: the function scope is extended to all subtypes of its argument.

Polymorphism can be static, i.e., resolved at compile time, or dynamic, i.e., resolved at runtime. It can apply to a single entity e.g., one of the arguments of a function, or multiple entities e.g., all the arguments of a function.  In a function call, the mechanism that selects the implementation to execute according the passed argument types is called dispatch.

In the following of this subsection, we will compare polymorphism provided by Julia, \Cpp and Python. Code examples illustrating our statements can be found in Appendix~\ref{appendix}.

Polymorphism is provided in \Cpp by two paradigms: one based on class inheritance, function overloading, and function overriding; the other based on templates. The first provides ad-hoc and subtype polymorphisms over functions, while the second provides universal and ad-hoc polymorphisms over both functions and types. By exploiting the {\tt concepts} feature introduced by \Cpp20, subtyping polymorphism support can be added to the templates. The functionalities of the two paradigms overlap.

In \Cpp, class non-static member functions take a special argument, the class instance reference ({\tt x}) or pointer ({\tt ptr}), through a dedicated syntax, {\tt x.f()} and {\tt ptr->f()}. Both static and dynamic polymorphism are supported over this argument, while only static polymorphism is provided for the other arguments. Object copy with implicit type conversion can make difficult to follow the polymorphism flow of an object.

Static ad-hoc polymorphism is supported over the arguments of global functions and static member functions. \Cpp class templates provide static universal polymorphism. One notable usage is the containers of the standard template library.

The class inheritance is twofold, it provides inheritance of the interface through the subtype polymorphism previously described and inheritance of the data fields, a type is an aggregation of its fields and all the fields of its supertypes. The bounding of the two inheritances can result in breaking encapsulation~\cite{10.1145/960112.28702} and it is often advocated to prefer composition to inheritance for the fields, as dictated by the ``second principle of object-oriented design'' of Ref.~\cite{GammaHelmEtAl95}.

Python provides single dynamic dispatch for the class instance argument of member functions.  It does not provide polymorphism for other arguments. Multiple dispatch emulation can be implemented in the function using conditions on the argument types or using a decorator~\cite{py_multidispatch}.

Julia provides ad-hoc and universal polymorphism, including subtype polymorphism~\cite{10.1145/3276483}, within a consistent multiple-dispatch paradigm. Extending the dynamic dispatch of \Cpp to all arguments of a functions makes it extremely powerful, especially in terms of code re-usability. The Julia multiple dispatch exploits JIT compilation and the classification into static and dynamic dispatches is less relevant here: a specialized function is compiled only before its use, although the behavior is always consistent with dynamic dispatch; inlining and other compile-time optimizations can be performed despite the dynamic behavior. Nevetheless, this optimisation is subject to the ability of the compiler to infer the type of the passed arguments and requires some attention from the developer. In particular, when code performance is important, the developer must make sure that the return type of a function can be inferred from the types of the its arguments. Julia ad-hoc and universal polymorphism uses a simple syntax similar to function overloading, but with argument types specified only when required, either to extend a function or to enforce the types of arguments.

The implementation of a two-argument function to be called by default, in absence of a more specialized implementation fitting better with the types of the passed arguments, will be defined as \mintinline{julia}{function(x, y)...end}. Its specialization for a first argument of type A or of a subtype of A will be defined by suffixing the first argument \mintinline{julia}{x} with \mintinline{julia}{::A}. It can be further specialized for a first argument of type A or a subtype of A and second argument of type B or of a subtype B, by annotating both arguments respectively with \mintinline{julia}{::A} and \mintinline{julia}{::B}, which will read as, \mintinline{julia}{function f(x::A, y::B)}.

Universal polymorphism for type definition is supported by parametric types. In the following example the type {\tt Point} has two fields of the same type, that must be a subtype of the {\tt Number} type.

\vspace{0.8em}
\begin{minted}{julia}
struct Point{T <:Number}
  x ::T
  y ::T
end
\end{minted}

Contrary to \Cpp, in Julia, subtyping does not involve field inheritance. Data aggregation must be done using composition, enforcing the ``second principle of object-oriented design''. Subtypes are used to define a type hierarchy for the subtype polymorphism. The hierarchy tree is defined with abstract types that do not contain data, only the leaves of the tree can be a concrete type.

Because variable assignment and function argument passing do not trigger an object copy, Julia is not affected by the difficulty encountered in \Cpp mentioned before. This is demonstrated with the ``King of Savannah'' example discussed in the Appendix~\ref{appendix}.

\section{HEP computing requirements}

Because the program codes used in HEP research are very large, with high interdependence, a code typically uses many open source libraries developed by other authors; thus the effort to change programming language is consequential. The adoption of a new language can happen only if it brings a substantial advantage over the already used paradigm.

The key advantage of Julia that can make the language switch worthwhile is the simplification that will arise from using a single language in place of a combination of two, \Cpp and Python. 

HEP computing is wide and includes many use cases: automation for the controls of the experiment, data acquisition, phenomenology and physics event generation, simulation of the physical experiment, reconstruction of physics events\footnote{In HEP experiments, we observe collisions of subatomic particles or atoms. The result of a collision that produces new particles is called an event. Detectors, that can be complex apparatus as large as $46\,\text{m}\times25\,\text{m}$ producing tens of millions of MBytes of data per second, are used to capture the event.} from recorded data, analysis of the reconstructed events, and more. 

We will review in this section the properties required for event analysis, event reconstruction, event simulation and event generation. We will start with general features, common to all the use cases.

\subsection{General features}

\subsubsection{An easy language}

The easy language is one-side of the high-level and high-performance coin property that would motivate the adoption of Julia as a programming language. It is easy at least in two ways: easy, imperative syntax, and free of strong typing when writing code. The surface syntax of Julia largely resembles Python, MATLAB (control flow, literal array), while also getting inspiration, such as the do-block, from Lua and Ruby. It has all the high-levelness one would expect from a language: higher-order functions (functions can be returned and passed as variables), native N-dimensional arrays, nested irregular arrays (arrays of different-size arrays), and a syntax for broadcasting over arrays.

As a syntax comparison example, a for-loop will look like the following in Python, Julia, and \Cpp.


\vspace{0.8em}
\begin{minted}{python}
# Python
a = 0.
for i in range(1,11):
  a += i
a /= 10.
\end{minted}

\vspace{0.8em}
\begin{minted}{julia}
# Julia
a = 0.
for i in 1:10
  a += i
end
a /= 10.	
\end{minted}

\vspace{0.8em}
\begin{minted}{c++}
// C++
auto a = 0.;
for(auto i = 0; 
    i < 11.; ++i){
  a += i;
}
a /= 10.;
\end{minted}

\vspace{0.8em}

We will note in this example that Julia is free of type declaration, just like Python. In this example, we use the \Cpp \mintinline{c++}{auto} type declaration feature to achieve the same goal. It is worthy of mention that, as in \Cpp, Julia code interpretation is not sensitive to changes in indentation: appending two spaces at the beginning of the last line, will change the behavior of the Python code only.

Julia supports list comprehension, like Python, as illustrated in this example that creates a vector with the series $1, 1/2,\dots,1/10$:


\vspace{0.8em}
\begin{minted}{julia}
# Julia
v = [ 1/x for x in 1:10 ]
\end{minted}

\vspace{0.8em}
\begin{minted}{python}
# Python
v = [ 1.0/x for x in range(1, 11) ]
\end{minted}

\vspace{0.8em}

\codename{NumPy}~\cite{numpy} function vectorization is provided natively in Julia for all functions and operators, including those defined by the user, through the broadcast operator: a dot prefix is used to specify that the function must be applied to each element of a vector or of an array. The syntax is illustrated below.


\vspace{0.8em}
\begin{minted}{julia}
# Julia
v = [1, 1, 1] ./ [1, 2, 3]
\end{minted}

\vspace{0.8em}

\begin{minted}{python}
# Python
import numpy as np
v =  np.array([1, 1, 1]) \
   / np.array([1, 2, 3]) 
\end{minted}

\vspace{0.8em}

The following example illustrates the native support of linear algebra and multi-dimensional arrays and highlight the concise syntax it provides. It solves the simple equation,
\begin{equation*}
\begin{pmatrix}
	1 & 1 \\
 	1 & -1	
 \end{pmatrix} x = \begin{pmatrix}
 2\\ 0
\end{pmatrix}
\end{equation*}.


\begin{minted}{julia}
# Julia
m = [ 1 1; 1 -1 ] \ [ 2, 0 ]
\end{minted}

\vspace{0.8em}

\begin{minted}{python}
# Python
import numpy as np
m =  np.linalg.solve(
       np.array([[1, 1], [1, -1]]),
       np.array([2, 0]))
\end{minted}

\vspace{0.8em}

Because broadcasting has its own syntax, Julia is able to use mathematical operators ``correctly'' when they are not broadcast, instead of relying special names e.g., matrix multiplication ({\tt np.matmul}) and exponentiation ({\tt np.ling.expm}). 

Two additional language design choices are worth noting for their contribution to make the language easy to use without sacrificing performance. The first is the evaluation of the function default parameter values, done at each call, instead of once for all in Python. Thanks to this choice, a function \mintinline{python}{f(x, v=[])} that appends the element x to vector v and returns the latter will always return a one-element when called as {\tt f(x)} in Julia, while it will return a vector growing in size at each new call in Python. The second is the copy performed by the updating operators ({\tt+=}, {\tt *=},..) instead of the in-place operation done by Python. In the Julia code, \mintinline[breaklines]{julia}{A = [0, 1]; B = A; B += [1, 1]}, the operator {\tt +=} will not modify the content of the vector {\tt A}, it is syntactically strictly equivalent to {\tt B = B + [1, 1]}), while it will in the Python code \mintinline[breaklines]{python}{A = np.array([0, 1]); B = A; B += np.array([1, 1])}. We judge these two Julia behaviors more natural and more likely to match to what a non-expert would expect when writing or reading the code.
 
As we illustrated with few examples, the Julia language is as easy as Python and sometimes easier thanks to a native support of features provided by external packages in Python that allows for a more concise and natural syntax.

\subsubsection{Performance}
\label{perf_comp}

\begin{figure}
  \centering
  \includegraphics[width=\linewidth]{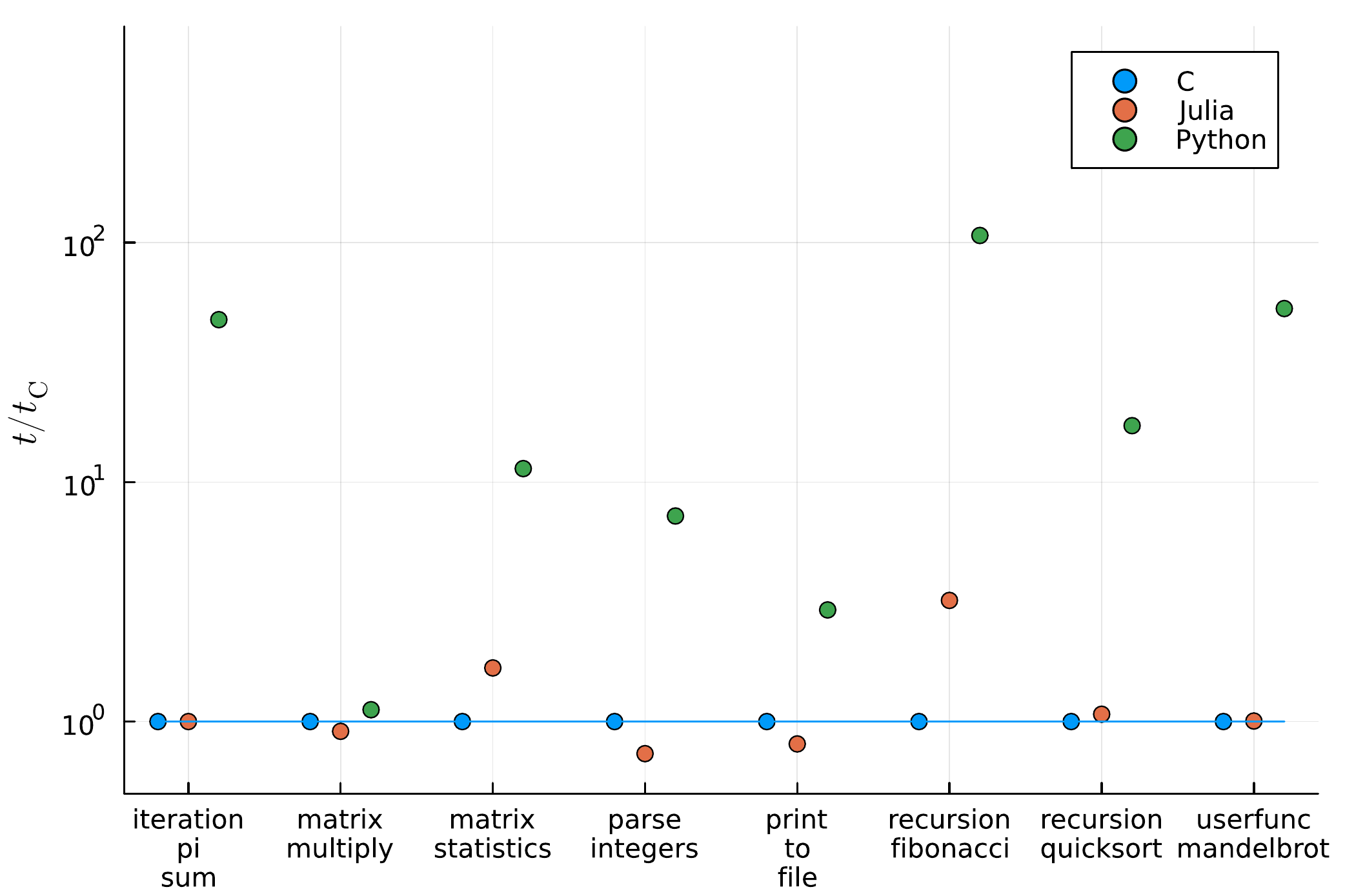}
  \caption{Comparison of C/\Cpp, Python and Julia language performance for a set of short algorithms. Open BLAS, together with \codename{NumPy} in the Python case are used for matrix operation. The score is defined as the time to run the algorithm divided by the time to run the C version of the same algorithm.}
  \label{fig:microbench}
\end{figure}

The other major advantage to Julia is high performance. Julia provides performance similar to \Cpp, and in some cases even surpassing \Cpp, as can be seen in Fig~\ref{fig:microbench}. The shown comparison is obtained by repeating the microbenchmark\footnote{\url{https://julialang.org/benchmarks/}} for Julia version~1.9.0rc1 and Python~3.9.2. C and Julia implementations use Open BLAS for the matrix operations, while Python uses \codename{NumPy} (version 1.24.2) and Open BLAS. This benchmark compares the run time of some short algorithms implemented in a similar way in the different languages.  The results are divided by the time the C/\Cpp implementation takes. The GNU compiler \codename{gcc} version 10.2.1 has been used. The test is performed on a laptop equipped with a 11th Gen Intel(R) Core(TM) i5-1135G7$@$2.40GHz CPU and 16$\,$GB of random access memory (RAM) running the Linux operating system compile for x86 64-bit architecture. The 64-bit flavor Open BLAS library version 0.3.21 is used. This setup is used for all performance tests described in this paper, if not specified otherwise. The score goes from 0.73 to 1.67 (smaller is better) for Julia and 1.12 to 107 for Python. C is doing the best with respect to the two other languages for the recursive Fibonacci algorithm, implemented in Julia as below.


\vspace{0.8em}
\begin{minted}{julia}
fib(n) = n < 2 ? n : fib(n-1) + fib(n-2)
\end{minted}
\vspace{0.8em}

 This benchmark tests the performance for recursive calls. While expert developers typically avoid it for performance reasons, a recursive expression is the easiest and most natural way to implement a recursive algorithm. The mathematical series $u_{n+1}=f(u_n)$ maps directly to a recursive computing function call. We compute the 20th Fibonacci series elements, which results in 21,891 nested calls, a good example of recursive calls. The C/\Cpp implementation is doing better because of a tail recursion optimization performed by the compiler, that removes one of the two nested calls, disabling this optimization leads to performance a little worse than with Julia. The gain from this optimization is far from the one obtained by using a for-loop implementation instead of the recursion. Such implementation runs $\approx$1000 times faster. The tail recursion optimization does not work for the recursive quicksort, leading to similar performance from C/\Cpp and Julia (7\% difference in favor of C/\Cpp).

We can use LHC open data to test performance on HEP-oriented code. We make this test with a di-muon analysis on CMS data of LHC Run-1, from 2011 and 2012. The analysis consists of measuring the spectrum of the mass of the system made of a muon and an antimuon, produced in proton-proton collisions at the center-of-mass energy $\sqrt{s} = 7\,$TeV.  It uses data in which the muons and antimuons are already reconstructed and identified. It does not correct for instrumental efficiencies, contrary to the published physics results.

Different implementations have been compared: the for-loop based Julia implementation from Ref.~\cite{bib:cms_dimu_jl}, the equivalent for-loop based implementations in Python and \Cpp, the \codename{ROOT} data frame (RDataFrame)  implementation from Ref.~\cite{bib:cms_dimu_pycxx}, its equivalent in \Cpp in two flavors, and a data frame based implementation done in Julia using the \codename{DataFrames.jl} package~\cite{bib:dataframes}. In the data frame implementation, the table rows are first filtered to obtain a data frame with the di-muon events of interest, then a column with the dimuon mass is added to the data frame, and finally a histogram is filled. RDataFrames use lazy operations, and only the histogram is materialized, limiting the memory footprint. In the first flavor of the \Cpp implementation, the formula to compute the mass is provided as a character string and the code for this computation is compiled JIT. In the second flavor a user-defined \Cpp function is provided to compute this mass.

The input data are read from a file stored in the \codename{ROOT} format with compression turned off. The \codename{UnROOT.jl} package~\cite{bib:UnROOT} (version~0.9.2) is used to read the file with the Julia code. This package is written in pure Julia. The native ROOT library (version~6.26/10) serves to read the files from \Cpp and Python. The GNU \codename{gcc} compiler (version Debian-10.2.1-6) is used with a level-three optimization (option {\tt -O3}). When JIT compilation is involved (the cases of Julia and JIT RDataFrame) the event analysis function is first run on a ten-event data file to trigger compilations before performing the timing on 1 billion events. For the Julia implementations, subsequent compilations occur during the timing loop; they represent only 1.1\% of the time. In the case of JIT RDataframe an overhead (time independent of number of processed events) of $5.0\pm0.2\,$ms ($\Cpp$ version) or $11.2\pm2\,$ms (Python version) is present in spite of the warm-up. The overheard is subtracted from the measurement. The obtained numbers are provided in Table~\ref{tab:dimu_time}. We observe that slight changes of source code can change the runtime of the \Cpp for-loop and native RDataFrame implementations beyond the statistical uncertainties. This effect is estimated by varying the code outside if the timed loop (addition of a print-out statement, change of code statement order) and included in the quoted uncertainties. For the other implementations, no significant change is observed and the quoted uncertainty include the statistical component only (at 68\% confidence level).

\begin{figure}
    \centering
    \includegraphics[width=1.\linewidth]{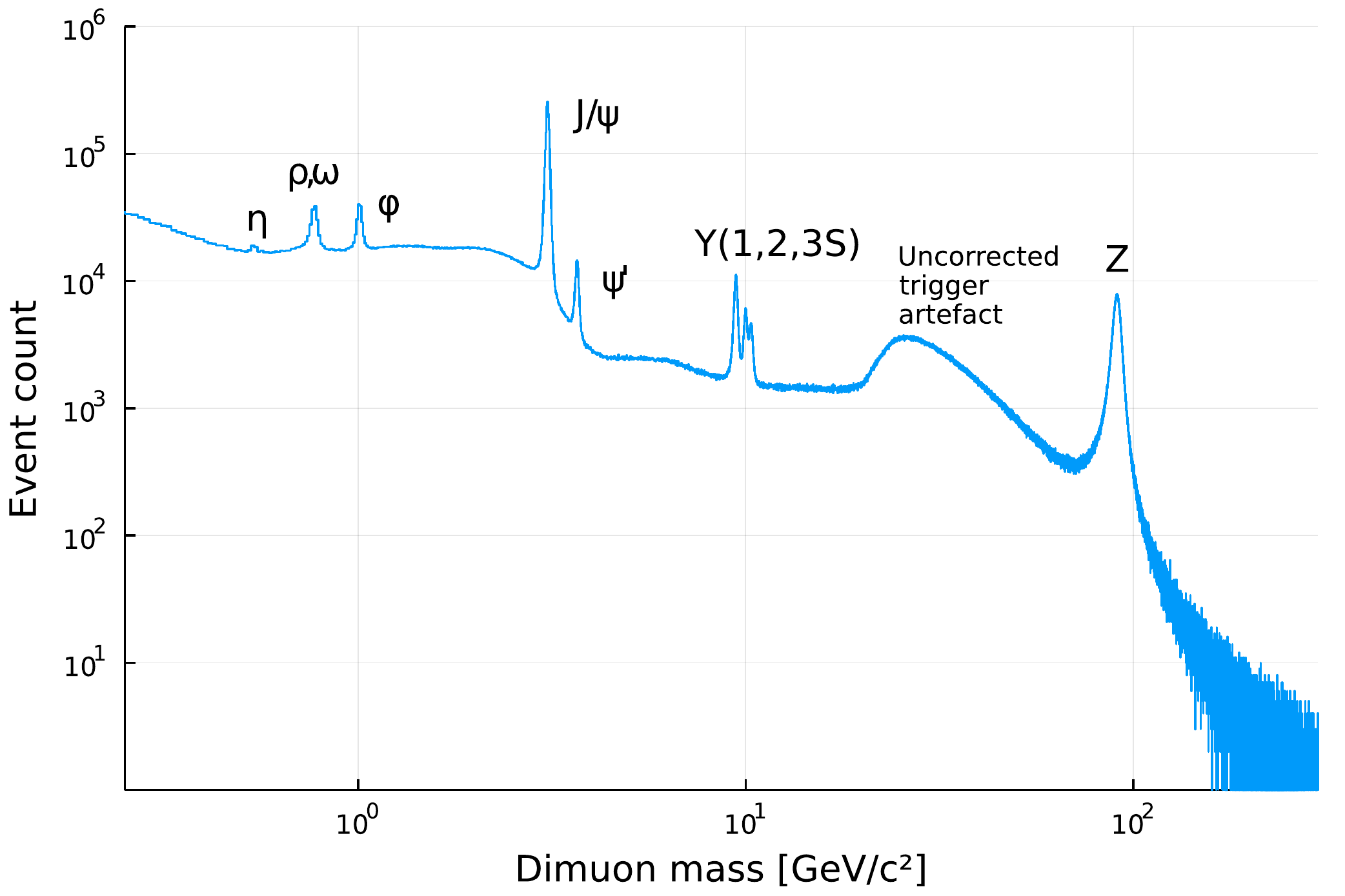}
    \caption{Dimuon spectrum obtained from the CMS open data of Run 2012 with the Julia implementation of the analysis.}
    \label{fig:dimuon}
\end{figure}

In this example, the for-loop Julia implementation runs the fastest, the \Cpp for-loop implementation is slightly behind (11\% slower). The Julia implementation using data frames takes 21\% less time to run than with \Cpp RDataFrame. The Python for-loop implementation is 1000 times slower than with Julia. Delegating the loop to an underlying compiled library (in our case the \codename{ROOT} library) is not sufficient to achieve good performance with Python: the RDataFrame python implementation is 2.2 (resp. 2.8) times slower than the Julia data frame (resp. for-loop) implementation. The \Cpp RDataFrame implementations are slower than the Julia and \Cpp for-loop implementations by a factor from 1.4 to 2.1 depending on the implementations we compare. The dimuon spectrum obtained with the Julia code is shown in Fig.~\ref{fig:dimuon}.

The data frame benchmark includes the insertion of a column in the data frame with the dimuon mass. In the Julia case, the insertion is not needed for the analysis itself, but keeping it is interesting for benchmark purpose. The data frame returned by \codename{UnROOT} does not allow direct insertion and the selected rows are copied to a DataFrames.jl data frame supporting such an insertion. That leaves room for improvements; we estimate that improved tools that would allow such insertion with no copy would reduce the runtime by 16\%.

For Python, the pure python library \codename{Uproot}~\cite{bib:uproot} can be used instead of the native \codename{ROOT} library to read the data. This library loads all the data of a file into the memory, similar to the Julia data frame implementation. The data can be provided as a set of Awkward Arrays~\cite{bib:awkwardarrays}, \codename{NumPy} arrays, or as a Pandas data frame~\cite{bib:pandas}. All these data structures support vectorized operations permitting a delegation of the event loop to underlying compiled libraries improving the running time. The results are shown in Table~\ref{tab:dimu-uproot}. The measurement is done with \codename{Uproot} version 4.3.4 (with \codename{awkward} package version 1.10.3). The implementation using Awkward Arrays operating on a vector of all events runs faster than the Python RDataFrame implementation and is only 1.6 times slower than with the Julia for-loop. We note that the Python's performance is highly dependent on the algorithm implementation: the time ratio with respect to the Julia for-loop goes up to 63 for a vectorized implementation using Pandas data frames and to 1200 with the event loop.

Running on a 61.5 million event file shows that the for-loop and RDataFrame implementations scale well with larger input files with no penalty on the event throughput as we could have expected. The other implementations would require modifications in the code in order to process events in chunks and reduce the memory usage. The awkward array implementation requires 14.5$\,$GiB, at the limit of 15$\,$GiB available on the machine used for the measurement, while the Pandas and Julia data frame versions exceed this limit.

We see in this example that Julia is performing similar or better than \Cpp frameworks. For an event loop, Python is slower by three-orders-of-magnitude than Julia. Vectorization of event processing serves as a mitigation of Python's slowness by delegating the event loop to underlying compiled libraries and sacrifice flexibility, without achieving the performance of \Cpp and Julia\footnote{Recently, it became possible to use Numba + Awkward Array to enable fast loops, sacrificing some Python features due to the more strict compiling model}.

\begin{table}
  \caption{Comparison of the runtime of the di-muon spectrum analysis for implementations performed in different programming languages. The time corresponds to a run over 1 million events.} 
  \label{tab:dimu_time}
  \centering
  \begin{tabularx}{\linewidth}{Xr@{$\,$}r@{}l}
  \toprule
  Implementation     & \multicolumn{3}{c}{\makecell{Time to process\\$10^6$ events}}\\
  \hline
  Julia for-loop          & 0.147   &$\pm$&0.0014$\,$s\\
  Julia Dataframe         & 0.1839  &$\pm$&0.0019$\,$s\\
  Python for-loop         & 153.7   &$\pm$&5.7$\,$s \\
  Python RDataFrame       & 0.4083  &$\pm$&0.0083$\,$s \\ 
  \Cpp for-loop           & 0.1627  &$\pm$&0.0019$\,$s\\
  \Cpp RDataFrame         & 0.2338  &$\pm$&0.00031$\,$s\\
  \Cpp RDataFrame JIT     & 0.3051  &$\pm$&0.0023 $\,$s\\
  \bottomrule
\end{tabularx}

\end{table}

\begin{table}
  \caption{Runtime of the dimuon spectrum analysis for three Python implementations  using the \codename{Uproot} library to read the data.}
  \label{tab:dimu-uproot}
  \begin{tabularx}{\linewidth}{Xr@{$\,$}r@{}l}
  \toprule
  Implementation & \multicolumn{3}{c}{\makecell{Time to process\\$10^6$ events}}\\
  \hline
  Vectorized with Awkward Arrays   & 0.2343 &$\pm$&$0.0027\,$s\\
  Vectorized with Panda dataframes & 9.225 &$\pm$&$0.081\,$s\\
  For loop                         & 177.2 &$\pm$&$1.8\,$s \\
  \bottomrule
\end{tabularx}

\end{table}

\subsubsection{Interoperability with legacy code}\label{sec:interoperability}

HEP computing is based on a heritage of program code written over decades. Interfacing to libraries developed in \Cpp and Fortran is unavoidable, apart from the last-step of analysis domain (and even here it would still be an attractive feature). Julia can natively call C and Fortran functions with no overhead compared to calling them from their native language. Examples of such calls are given in Listings~\ref{lst:fortran-from-julia} and~\ref{lst:c-from-julia}. For convenience, a wrapper function written in Julia can be used to handle errors, as in the example in Listing~\ref{lst:c-err-handling}.

\begin{figure}
    \centering
	\includegraphics[width=\linewidth]{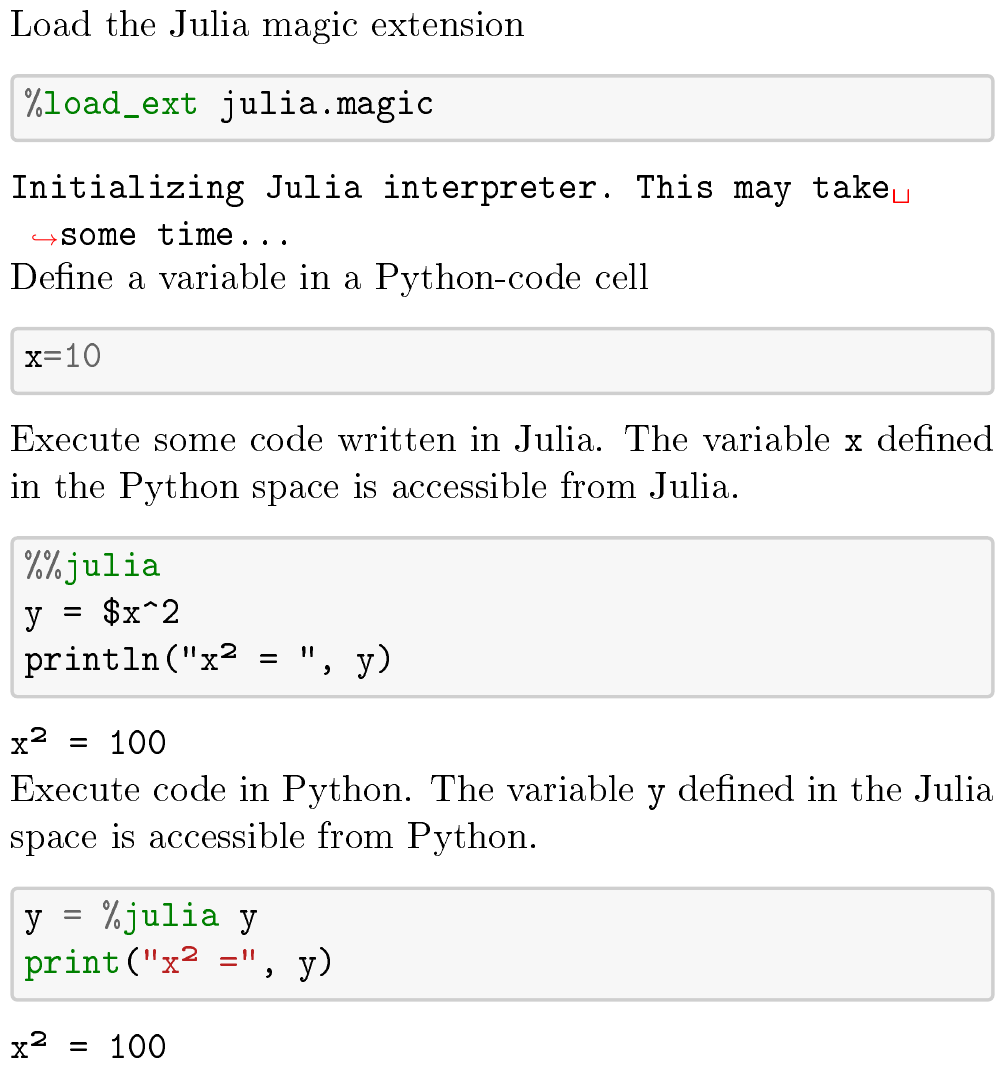}
	\caption{Example of a Jupyter notebook mixing cells with Julia and Python code.}
	\label{fig:nb-julia-magic}
\end{figure}

Bindings to Python are supported thanks to the \codename{PyCall} package. The interface is very convenient and transparent in both directions, Python from Julia and Julia from Python, as we can see in the examples provided in Listings~\ref{lst:python-from-julia} and~\ref{lst:julia-from-python}. In a Jupyter notebook, in addition to calling a Julia function from a notebook running a Python kernel using these interfaces and vice-versa, it is possible to write Julia code in cells of a notebook using a Python Kernel, and mix cells written in Julia and in Python languages, as illustrated in Fig.~\ref{fig:nb-julia-magic}.

\begin{listing}
    \begin{minted}{julia}
function compute_dot(DX::Vector{Float64}, 
                     DY::Vector{Float64})
  @assert length(DX) == length(DY)
  n = length(DX)
  incx = incy = 1
  product = ccall((:ddot_, "libLAPACK"),
              Float64,
              (Ref{Int32}, Ptr{Float64},
              Ref{Int32}, Ptr{Float64}, 
              Ref{Int32}),
              n, DX, incx, DY, incy)
  return product
end
\end{minted}
  \vspace{0.8em}
  \caption{Example of call from Julia of a function implemented in  Fortran~\cite{bib:juliamanual}}
  \label{lst:fortran-from-julia}
\end{listing}

\begin{listing}
    \begin{minted}{julia}
path = ccall(:getenv, Cstring, (Cstring,), 
             "SHELL")
println(unsafe_string(path))
\end{minted}
    \vspace{0.8em}
    \caption{Example of call of a function of a C-library from Julia~\cite{bib:juliamanual}}
    \label{lst:c-from-julia}
\end{listing}

\begin{listing}
    \begin{minted}{julia}
function getenv(var::AbstractString)
  val = ccall(:getenv, Cstring, 
              (Cstring,), var)
  if val == C_NULL
    error("getenv: undefined variable: ", 
          var)
  end
  return unsafe_string(val)
end
    \end{minted}
  \vspace{0.8em}
  \caption{Example of a wrapper in Julia to handle errors from a c-library function~\cite{bib:juliamanual}}
    \label{lst:c-err-handling}
\end{listing}

\begin{listing}
    \begin{minted}{julia}
# Enable Python call:
using PyCall

# Inport a python module:
math = pyimport("math")

# Use it as a Julia module:
math.sin(math.pi / 4)
\end{minted}
    \vspace{0.8em}
    \caption{Python function can be called transparently from Julia. Example of call of a function from the Python \codename{math} package.~\cite{bib:juliamanual}}
    \label{lst:python-from-julia}
\end{listing}

\begin{listing}
    \begin{minted}{python}
$ python3
>>> import julia
>>> julia.install()
>>> from julia import Base
>>> Base.sind(90)
1.0
\end{minted}
  \vspace{0.8em}
    \caption{The Python package named \codename{julia} allows use of Julia packages within Python codes.}
    \label{lst:julia-from-python}
\end{listing}

The \codename{CxxWrap} package~\cite{bib:cxxwrap} can be used to add Julia bindings to \Cpp libraries. Once bound, the library is accessed transparently from Julia as if it was a native Julia package. The \mintinline{c++}{object.method(args...)} and \mintinline{c++}{object_ptr->method(args...)} \Cpp like method calls translate into the Julia-like call \mintinline{julia}{method(object, args...)}. The package philosophy is similar to \codename{Boost.Python}~\cite{bib:boost-python} and \codename{Pybind11}~\cite{bib:pybind11}: the bindings are produced with few lines of \Cpp code, one line per class and one line per method, which must be compiled as a shared library. The package provides all the flexibility to expose a different Julia interface to the \Cpp one, for instance to adapt it to the Julia context and style of programming. \codename{CxxWrap} internally uses the built-in Julia-C interface, used for the interface between shared libraries and Julia.
The \Cpp standard template library \mintinline{c++}{vectors} and \mintinline{c++}{val_arrays} are mapped to Julia \mintinline{julia}{Vectors} with zero copy.

The \codename{WrapIt} project~\cite{bib:wrapit} has demonstrated that binding code can be generated automatically from a library's header files, which would make the process of adding Julia bindings to \Cpp libraries very easy. Automation of this Julia binding has been tested on the \codename{ROOT} libraries, and we have been able to produce, draw, and fit histograms and graphs (\mintinline{c++}{TGraph} class). The fit has been tested with both functions defined in \codename{ROOT} and functions defined in Julia, demonstrating a perfect integration.

Unlike direct calls into C or Fortran libraries via the \mintinline{julia}{ccall} function, calls between \Cpp and Julia have to go through the intermediate layer created by the wrapper code. We perform several measurements to estimate the overhead from the \Cpp-Julia interface. The measurement is performed on a call to the \codename{ROOT} \mintinline{c++}{TH1D::Fill} method, that adds a value to a histogram: we time a loop of 1 million calls and average the result to get the time per call. First, we create a shared library, that exports C functions, we call from Julia  with the \mintinline{julia}{ccall} method. The pointer to the histogram object is passed to the C function as a \mintinline{c++}{void*} type. When compared with a direct call to the \mintinline{c++}{Fill} method within the same \Cpp code, the call from Julia shows an overhead of 0.23$\,$ns. This overhead is unexpectedly smaller than when calling the wrapper function of the shared library from a program written in C: observed overhead of 0.74$\,$ns in this case. In the end, the call from Julia takes only 4\% more time than a direct call from \Cpp. It is 38 times faster than a call from Python. Measurement is also done for a binding based on \codename{CxxWrap}. All results are shown in Table~\ref{tab:cxxcall}.

We could imagine the Julia engine performing just-in-time compilation of \Cpp using the LLVM infrastructure it uses for the Julia code. The \codename{Cxx} package~\cite{bib:cxx} is providing this feature for Julia releases from 1.1.x to 1.3.x. With this package we can access to a \Cpp library without the need of a \Cpp wrapper. Nevertheless, a Julia wrapper is needed to provide the same transparency---calls to the \Cpp functions similar as calls to a Julia functions.  Using this package the call to the \mintinline{c++}{Fill} function in our example is found to be as fast as when using the C interface, as shown in  Table~\ref{tab:cxxcall}. We used Julia version 1.3.1 to perform this measurement. The \codename{Cxx} approach is a good alternative to \codename{CxxWrap}.

In Ref.~\cite{Stanitzki:2020bnx}, \codename{CxxWrap} was used to interface to the \codename{LCIO} \Cpp library~\cite{bib:lcio} to read ILC~\cite{Behnke:2013xla} simulated events and to \codename{Fastjet}~\cite{Cacciari:2011ma,Cacciari:2005hq} to cluster hadronic jets. The loss of event throughput compared to a code uniformly written in
\Cpp was 37\%. 

\begin{table}
  \caption{Mean time to call the \mintinline{c++}{Fill} method of a \codename{ROOT} histogram from \Cpp, Julia and Python. The time corresponding to a single call is averaged on $10^{6}$ \mintinline{c++}{h->Fill(1.)} calls. Three cases are considered for Julia: use of the plain Julia C interface (``C API''), use of \codename{CxxWrap}, and use of \codename{Cxx}. For reference the time to call the same function from C/\Cpp, within the same code (``\Cpp'') and through the shared library developed the Julia C interface, is also measured.}
  \label{tab:cxxcall}
\begin{tabularx}{\linewidth}{Xr@{$\,$}r@{}l}
  \toprule
  & \multicolumn{3}{c}{Mean time} \\
  & \multicolumn{3}{c}{[ns]} \\
  \hline                                       
  \Cpp                        & 5.74 &$\pm$&0.01\\
  C API from C                & 6.48 &$\pm$&0.04\\
  C API from Julia            & 5.97 &$\pm$&0.03\\
  Julia - \codename{CxxWrap}  & 8.21 &$\pm$&0.04\\
  Julia - \codename{Cxx}      & 5.97 &$\pm$&$(<0.01)$\\
  Python                      & 226  &$\pm$&5  \\
  \bottomrule
\end{tabularx}

\end{table}

\subsubsection{Support of standard HEP data formats}

Different file formats are used to store HEP data and supporting them is crucial to a streamlined physics analysis experience.

The file formats currently used to store physics events are mainly \codename{HepMC}, \codename{LHE}~\cite{bib:LHE}, \codename{LCIO}, and \codename{ROOT}. The pachages \juliapkg{LHE.jl}~\cite{bib:LHEjl} and \juliapkg{LCIO.jl}~\cite{jan_strube_2021_4560484} provide supports for \codename{LHE} and \juliapkg{LCIO}. Two packages are available to read \codename{ROOT} files: \juliapkg{UpROOT.jl}~\cite{bib:uprootjl} and  \juliapkg{UnROOT.jl}~\cite{bib:UnROOT}. 

The \codename{UpROOT.jl} package uses the \codename{Uproot} pure-Python library to provide read and write support. When using this package, a loop on events typically suffers of the same performance penalty as with Python. This has motivated the development of \juliapkg{UnROOT.jl}, a package written in pure Julia that provides a fast processing of events, as demonstrated in the performance measurements done in the previous section, which used this package. It leaves the flexibility to use an explicit event loop, with a small memory footprint, or works on vector of event quantities (``columnar analysis''). An event loop will look like the following code snippet, where \mintinline{julia}{Muon_pt} is a vector (transverse momenta of the muons contained in the event).
\vspace{0.8em}
\begin{minted}{julia}
for event in mytree
  # Access to a single-event quantity
  event.Muon_pt
end
\end{minted}
\vspace{0.8em}
A columnar analysis will look like the following.

\vspace{0.8em}
\begin{minted}{julia}
# Access to a vector of event quantities
# (themselves vectors of numbers)
mytree.Muon_pt
\end{minted}
\vspace{0.8em}

\juliapkg{UnROOT.jl} uses thread-local storage to maximize performance and maintain thread-safety. An event loop can be parallelized in several threads with little effort using the standard Julia \mintinline{julia}{@threads} macro:

\vspace{0.8em}
\begin{minted}{julia}
julia> @threads for event in mytree
# ... Operate on event
end	
\end{minted}
\vspace{0.8em}

The performance measurement presented in the ``\nameref{perf_comp}'' section are done in single-thread mode. There are limitations. First, this package does not support data write. Both \juliapkg{UpRoot.jl} and \juliapkg{UnROOT.jl} can access only to objects of a limited set of types, either stored as such in the file or in a \mintinline{c++}{TTree}. The supported types covers already a large set of use cases, but not schemes where data is stored as object of serialized \Cpp classes. Using the genuine \codename{ROOT} library via a Julia binding based on \codename{CxxWrap} can be an alternative approach when required. We have successfully read and write histograms (\mintinline{c++}{TH1} objects) and graphs (\mintinline{c++}{TGraph} objects) using this approach.

We expect the implementation of the support of \mintinline{c++}{RNTuple} to be easier than  \mintinline{c++}{TTree} that it is expected to replace, thanks to its design. Data are stored in column of fundamental types (\texttt{float}, \texttt{int},...)~\cite{Blomer:2020usr}, similar to Apache Arrow\cite{bib:apache_arrow}, which should ease support from programming languages other than \Cpp like Julia.

In the neutrino physics community, the industry-standard \codename{HDF5} and \codename{Parquet} have been used at scale, and these files can be readily read and write from Julia via their respective packages.

\subsubsection{Parallel computing}

Apart from having memory shared multi-threading, Julia also ships with out-of-core distributed computing capability as a standard library (\juliapkg{Distributed}). In fact, it is as easy to command an array of heterogeneous nodes in real-time via packages such as \juliapkg{ClusterManagers.jl}~\cite{bib:clustermanagers}, which can mimic \codename{Dask}'s experience~\cite{bib:dask} with a fraction of the code. For more advanced features, such as building out-of-core computation graphs, \juliapkg{Dagger.jl}~\cite{bib:dagger} provides facilities.

While these libaries allow distribution of execution within the Julia code, parallelization can be also done,  as with \Cpp and \codename{Python}, by running parallel jobs of the same executable using commands of a batch processing system, like HTCondor~\cite{bib:condor,bib:htcondor}, typically used in computer cluster facilities.

\subsubsection{Platform supports and license}
Julia is supported on all major platforms, a list of which can be found on the Julia website\footnote{\url{https://julialang.org/downloads/\#supported_platforms}}. Three different support tiers are provided. The platforms with full-fledged support (classified as tier 1) are, as of Oct, 2022:
\begin{itemize}
    \item macOS x86-64
    \item Windows x86-64 and CUDA
    \item Linux x86-64 and CUDA
    \item Linux i686
\end{itemize}

Worth noting that many platforms well on their way into tier 1, such as macOS with ARMv8 (M-series chips).

Julia is distributed under the MIT license along with vast majority of the ecosystem, which guarantees free use, modification, and re-distribution for any use case.

\subsubsection{Reproducibility}
\label{sec:reproducibility}

Julia includes a package manager and a general registry used by the whole community in an organized manner. In particular, each package contains a \fname{Project.toml} file, that records the dependency and compatibility with other packages in an uniform way. 

Furthermore, any binary dependencies are also captured by the package system: they are distributed as ``Artifact''---packages with names ending \codename{\_jll}---but still behave as normal packages when it comes to dependency and compatibility resolution. This eliminates a few problems, including running out of \codename{pip} space just because you depend on a large library (e.g., \codename{CUDA}). More details are giving in the ``\nameref{sec:pkging}'' section.

On the end-user side, one can easily capture the an environment by working with the \fname{Manifest.toml} file. While \fname{Project.toml} records compatibility and dependencies, Julia would try to use the latest compatible packages when instantiating the environment. \fname{Manifest.toml}, on the other hand, captures the exact versions of every package used (recursively) such that exact reproducibility can be guaranteed.

\subsubsection{Numerical optimization}
\label{sec:numericaloptim}

The statistical inference procedures relevant to HEP use numerical optimization heavily, from Maximal Likelihood Estimation (MLE) to scans over Parameters of Interest (POI) and obtaining the test statistics. Traditionally this is done by \codename{minuit2}~\cite{James:1975dr,bib:minuit2} in \codename{ROOT}, which uses the finite difference method to provide gradient information for some of its optimization.

Julia has a solid ecosystem in numerical optimization (\juliapkg{NLopt.jl}~\cite{bib:nlopts}, \juliapkg{Optim.jl}~\cite{bib:optimjl}, and meta algorithm package such as \juliapkg{Optimization.jl}~\cite{bib:optimizationjl} that brings local and global optimization together). Julia's performance has lead to most libraries being written in pure Julia, which means that optimization tasks can often use better algorithms such as Broyden–Fletcher–Goldfarb–Shanno (BFGS)~\cite{10.1093/imamat/6.1.76,10.1093/comjnl/13.3.317,goldfarb1970,shano1970} that rely on gradient provided by automatic differentiation. Support for automatic differentiation is further described in the ``\nameref{autodiff}'' section.

Construction of a complex probability distribution function is a common problem in HEP. Description of continuous spectra often requires a multicomponent probability density function (PDF) e.g., a sum of a signal component and a background component. In addition, the convolution with the model PDF with the experimental resolution is an essential for the HEP applications. The \codename{RooFit} framework is the standard tool for building complex high-dimensional parametric functions out of lower dimensional building blocks. As great convenience, the framework provides a homogeneous treatment of the PDF variables and parameters that can be fixed, restricted to a range, or constrained by a penalty to the likelihood functions. The framework is written in C++ and available in Python. A pure-python package \codename{zfit}~\cite{Eschle:2019jmu, Eschle:2020ghu} give an alternative solution to Python users that can better integrate with the scientific Python ecosystem.

Julia ecosystem offers a large variety of standard density functions in the \codename{Distributions.jl} package~\cite{JSSv098i16, Distributions.jl-2019}. The package largely exploits the properties of the standard density functions, such as moments and quantiles, which are computed using analytic expressions for the unbound PDFs. Moreover, flexible construction functionality is greatly missing. The mixture models of the \codename{Distributions.jl} are the holder for the multicomponent PDF, however, they cannot be used for fitting of the component fractions, the prior probabilities. Extension of the convolution functionality beyond a small set of low-level functions is required. The management of the distribution parameters is a key missing functionality in Julia modelling ecosystem.

\subsection{Specific needs for analysis of reconstructed events}

\subsubsection{Tools to produce histograms and publication-quality plot}
The statistics community in Julia has support for N-dimensional histograms with arbitrary binning in \juliapkg{StatsBase.jl}~\cite{bib:statsbase}, an extension to this basic histogram is implemented in \juliapkg{FHist.jl}~\cite{FHist}, which added support for bin error and under/overflow and for filling the histograms in an event loop, as typically done  in HEP analyses.

Many libraries of high quality are available for plotting from Julia. In the interests of standardization, the \juliapkg{Plots.jl}~\cite{bib:plots} package provides a front-end interface to many plotting packages, allowing easy switching from one to another. It supports the concept of recipes used by packages processing data to specify how to visualize them, without depending on the Plots package: the dependency is limited to the \juliapkg{RecipeBase.jl}~\cite{bib:RecipeBase_github} package which has less than 350 lines of code. The package supports, currently, 7 backends. It supports themes, which are sets of default attributes and provide a similar feature to the \codename{ROOT} \mintinline{c++}{TStyle} class. The back end selected by default in Plots is GR~\cite{bib:gr}, a rich visualization package providing both 2D and 3D plotting and supporting \LaTeX{} for text. The GR package, or its \juliapkg{GRUtils.jl}~\cite{bib:grutils} extension, can be used directly when a shorter warm-up time is needed before obtaining the first plot of a running session (see the ``\nameref{sec:latency}'' section for a discussion on the warm-up time). 

We should also mention the \juliapkg{Makie.jl} ecosystem~\cite{DanischKrumbiegel2021}, a rich plotting package targeting publication-quality plots, which is increasingly popular. This package supports the recipe and theme features, but is not itself supported by \juliapkg{Plots.jl}. For instance, the \juliapkg{FHist.jl} HEP-oriented histograming package mentioned before provides a recipe to plot the histograms. \juliapkg{Makie.jl} suffers from a longer time to obtain the first plot, even larger than with the \juliapkg{Plots.jl} package with its default backend \codename{OpenGL}.

Use of \LaTeX{} to generate high-quality plots has been popularized in HEP community with the plotting system of the Rivet Monte-Carlo event generator validation toolkit~\cite{Bierlich:2019rhm}. The PFGPlots~\cite{bib:pfgplots} and PFGPlotsX~\cite{bib:pfgplotsx} packages offer \LaTeX-based plotting. They are both supported by the \juliapkg{Plots.jl} package. The \juliapkg{Gaston.jl}~\cite{bib:gaston} package provides plotting using the popular \juliapkg{Gnuplot.jl} utility~\cite{bib:gnuplot}.

People used to the Python \mintinline{python}{matplot.pyplot} set of functions~\cite{bib:matplot} can use the \juliapkg{PyPlot.jl} package that provides a Julia API to this package. Those who prefer \codename{plotly} to \codename{matplot}, can use the \juliapkg{PlotlyJS.jl}, a Julia interface to \codename{plotly}. The high-level grammar of interactive graphics \codename{Vega-Lite}~\cite{bib:vegalite} is also supported, thanks to the \juliapkg{VegaLite.jl}~\cite{bib:vegalitejl} package that supports exports to bitmap and vector image files, including the PDF format, which is convenient to include in papers written with \LaTeX. Plotting can also be done on a text terminal, using the \juliapkg{UnicodePlots.jl}~\cite{bib:unicodeplots} package, supported by the Plot front end.

The visualization tool ecosystem for Julia is rich, with the added benefit of staying in the same environment as the analysis and enabling an interactive workflow.

\subsubsection{Notebook support}

A computational notebook is an interface for literate programming that allows embedding calculations within text. Notebooks have been made popular by Mathematica~\cite{bib:mathematica}, which has supported notebooks starting from its first version, 1.0, released in 1988. In HEP, notebooks are widely used by theoreticians for symbolic calculation e.g., with Mathematica, and by experimentalists, for data analysis, and plotting using Python or \Cpp as programming language.

The notebook system used with Python, Jupyter, fully supports Julia. The ``ju'' of {\bf Ju}pyter stands for Julia, while ``py'' stands for Python and ``er'' for the R language. The \codename{ROOT} analysis framework brings \Cpp support to Jupyter.

The notebook support for Julia is richer than for Python and C++. In addition to Jupyter,  \juliapkg{Pluto.jl}~\cite{fons_van_der_plas_2022_6916713} provides a new-generation notebook system for Julia. This system keeps track of the dependency of all calculations spread in the document and updates automatically any dependent results when a one of them is edited. Beyond being convenient, this automatic update provides {\em reproducibility}. 

\juliapkg{Pluto.jl} is also a very easy solution for interactive notebooks, where buttons, drop-down menus and slides can be included. This is useful for students. It can also be used to build a tool for experiments running shifters to analyze the data quality in quasi-realtime.

With \juliapkg{Pluto.jl}, notebooks are normal executable Julia files. Notebook functionality is offered through special comments. This helps with version control.

\subsection{Specific needs for physics event reconstruction, simulation and data acquisition trigger software}

Physics event reconstruction, simulation and trigger software are typically large codes developed by the experiment and project collaborations. The software stack of the LHC experiments is particularly large and complicated, due to the complexity of their detectors. The software is developed collaboratively by many developers, with different levels of software skills. Tools for both collaborative development and quality assurance are essential for all experiment software. Software distribution and release management are also important. The complexity of the \Cpp language, used in most of these frameworks, can limit the integration of contributions developed by students. This is more and more true given the growing use of high-level language (e.g., Python) as the teaching language for computing in universities, especially among natural science departments.

The Julia language and its ecosystem have been built using an open-source and community approach. Tools have been put in place and are widely adopted for efficient collaborative development. Julia comes with a standard and convenient package management system providing reproducibility, see the ``\nameref{sec:pkging}'' section. Julia has built-in unit testing, coverage measurement, and officially maintained continuous integration recipes and documentation generator. These are used in almost all of the Julia packages registered publicly, thanks to the streamlined experience and low barrier to entry.

The simulation software of the experiments depends on external libraries to simulate the underlying physics, such as Monte Carlo event generators, and on some others, like Geant4~\cite{AGOSTINELLI2003250}, to simulate the transport of the produced particles and their interaction with the detector. Interoperability with libraries written in C, \Cpp, or Fortran, as discussed in the ``\nameref{sec:interoperability}'' section, it is essential not to have to re-write all the external libraries in Julia.

Simulation and reconstruction is compute intensive and therefore good performance is essential: performance has a direct impact on the computing infrastructure cost. We have seen in the previous section that Julia meets the C/\Cpp performance and sometimes surpasses it. Code parallelization and efficient use of single instruction multiple data (SIMD) vectorization features of CPUs is essential at the LHC and for HL-LHC to efficiently use current hardware resources, with a high density of computing cores, including accelerators (e.g., GPU) that can count tens of thousands core~\cite{HEPSoftwareFoundation:2017ggl}. The Julia language provides a very good support for multi-threading: a loop can be parallelized by a single macro (\mintinline{julia}{@threads}), an operation can be made atomic by prefixing it with \mintinline{julia}{@atomic}, a more general lock mechanism is provided, asynchronous tasks, with distribution of tasks to different threads, is natively supported. Julia supports distributed computing, using its own communication mechanism but also using MPI~\cite{Byrne2021,bib:elemental_jl}. It is possible to use Julia's compiler to vectorize loops by using the \mintinline{julia}{@simd} macro or the more advanced \mintinline{julia}{@turbo} from the \juliapkg{LoopVectorization.jl} package~\cite{bib:loopvec}.

Due to its effective metaprogramming capabilities, Julia has great support for running code on heterogeneous architectures, Julia code can be compiled for Nvidia (CUDA), AMD (ROC) and Intel (oneAPI) GPUs via compiler written in Julia\footnote{\url{https://github.com/JuliaGPU/GPUCompiler.jl}}, without dependency on, for example, \Cpp CUDA or HIP library. Packages like \juliapkg{GPUArrays.jl} and \juliapkg{KernelAbstractions.jl} allow the use of exactly the same core algorithm written in Julia to be executed across different vendor platforms with minimal boilerplate code, which is a currently a unique feature among languages. 

On the more user-facing front, libraries such as \juliapkg{Tullio.jl}~\cite{bib:tullio} combine metaprogramming and kernel programming ability to allow users to express tensor operation with Einstein notations regardless of whether the array lives on RAM or GPU VRAM. This is very relevant for data preservation and for unifying effort to write algorithms once and run them everywhere.

The ability to run native Julia code on both CPUs and GPUs, combined with the support for automatic differentiation in Julia, makes Julia an excellent platform for machine learning (ML) research. This is especially true for advanced scientific machine learning that goes beyond combining conventional matrix-crunching ML-primitives/layers and uses physical/semantic models or mixes them with generic ML constructs.

\subsection{Specific needs for event generation and for phenomenology}

\subsubsection{Symbolic Calculations in Julia}

Julia is a fast, solid and reliable programming language with a well developed Computer Algebra System (CAS) such as \juliapkg{Symbolics.jl}~\cite{10.1145/3511528.3511535}, a language for symbolic calculations such as \juliapkg{Symata.jl}~\cite{symata}, and an interface to \codename{Mathematica} such as \juliapkg{MathLink.jl}~\cite{mathlink}, that could be widely used in HEP, considering the advantages Julia has.

\juliapkg{Symbolics.jl}~\cite{bib:symbolics} is a CAS written in pure Julia, which is developed by the SciML community~\cite{bib:sciml} who also maintain the state-of-the-art differential equations ecosystem~\cite{rackauckas2017differentialequations}. The package has scalable performance and integrates with the rest of Julia ecosystem thanks to its non-OOP design and multiple dispatch~\cite{10.1145/3511528.3511535}. Some of the main features of \juliapkg{Symbolics.jl} include pattern matching, simplification, substitution, logical and boolean expressions, symbolic equation solving, support for non-standard algebras with non-commutative symbols, automatic conversion of Julia code to symbolic code and generation of high performance and parallel functions from symbolic expressions~\cite{juliasymbolics}, which make it even more interesting for possible applications in HEP. At the heart of \juliapkg{Symbolics.jl}, we find \juliapkg{ModelingToolkit.jl}, a symbolic equation-based modeling system~\cite{ma2021modelingtoolkit}, and \juliapkg{SymbolicUtils.jl}, a rule-based rewrite system~\cite{symbolicutils}.

\juliapkg{Symata.jl}~\cite{symata} is a language for symbolic computations in which some features, such as evaluation, pattern matching and flow control, are written in Julia, and symbolic calculations are developed by wrapping \codename{Sympy}, a python library for symbolic mathematics.

\juliapkg{MathLink.jl}~\cite{mathlink} is a Julia language interface for the Wolfram Symbolic Transfer Protocol (WSTP) (this requires the installation of Mathematica or the free Wolfram Engine to run properly). The interface is a \texttt{W""} string macro used to define Mathematica symbols. \juliapkg{MathLinkExtras.jl}~\cite{mathlinkextras} adds extra functionalities such as \codename{W2Mstr}, which allows the conversion of Julia MathLink expressions into \codename{Mathematica} expressions, and \mintinline{julia}{W2Tex} which converts Julia \juliapkg{MathLink.jl} expressions into \LaTeX{} format. And, finally, one can evaluate the expression in \codename{Mathematica} using \mintinline{mathematica}{weval}.

\subsubsection{Event Generators}

To be prepared for future needs for event generation \cite{HSFPhysicsEventGeneratorWG:2020gxw}, it is conceivable to rewrite parts of the existing event
generators in Julia and making use of modern parallelisation
technologies. One of the most demanding tasks in event generation is
the evaluation of matrix elements and cross sections, where Julia
provides several useful tools.\\ The package \juliapkg{Dagger.jl} is a
framework for out-of-core and parallel computing written in pure Julia. It is
similar to the python library \codename{Dask} and provides a scheduler for parallelized execution of computing tasks represented as a directed acyclic graphs (DAGs). Such DAGs could be used to represent the evaluation of matrix elements in terms of elementary building blocks, similar to \codename{HELAS}-like functions in \codename{Madgraph4GPU} (see e.g., \cite{Valassi:2021ljk}). Furthermore, \juliapkg{Dagger.jl} supports the selection of different processors as well, making it possible to be use for distributed computing on GPU as well (see e.g., \juliapkg{DaggerGPU.jl}~\cite{bib:daggergpu}).

\subsection{Feature summary}

The Table~\ref{tab:features} summarizes the programming language and ecosystem features we have identified as required for HEP. It is surprising how Julia language manages to fulfill almost all of these features. We should note that the interoperability is less transparent for \Cpp than with the other languages as it requires to write a code wrapper. Many HEP file format are already supported, including for \codename{ROOT} files, without covering the full versatility offered by this format.

\begin{table}
  \caption{Summary of features needed for HEP applications and their availability in the Julia ecosystem.}
  \label{tab:features}
  {
\def\arraystretch{1.4}
\begin{tabularx}{\linewidth}{X|c}
  \toprule
Requirement & \makecell{Fulfilled\\by Julia}\\
\hline
Easy to learn for HEP physicists & ✓ \\
Performance & ✓ \\
Interoperability with legacy code  & ✓ \\
Support for HEP standard formats & partial \\
Support for architectures and open license & ✓ \\
Cross-platform reproducibility & ✓ \\
Tools to perform optimization/minimization & ✓ \\
Histogramming & ✓ \\
Plotting with publication quality  & ✓ \\
Notebook support & ✓ \\
Tooling for large project (unit tests, continuous integration, software distribution)  & ✓ \\
SIMD and multi-threading & ✓ \\
Distributed parallel computing & ✓ \\
Native GPU support  & ✓ \\
Machine learning libraries & ✓ \\
Computer Algebra System & ✓ \\
  \bottomrule
\end{tabularx}}

\end{table}

\section{The bonuses}
In addition to solving the two-language problem and the mandatory features listed in the previous section, the Julia ecosystem will provide other advantages over the \Cpp and Python languages currently used.

\subsection{Packaging}
\label{sec:pkging}

Julia comes with a built-in package manager,  \juliapkg{Pkg.jl}.  It builds on the experience of package managers in other ecosystems, and it can be used to install packages and manage ``environments'', similar to the concept of virtual environments in Python.  A Julia environment is defined by two files:
\begin{itemize}
\item \fname{Project.toml}: this file records version and UUID of the current project, it also contains the list of direct dependencies of this project, as well as the compatibility bounds with these packages and Julia itself. Moreover, all Julia packages follow semantic versioning (semVer~\cite{bib:semVer}): version numbers are composed of three parts, major, minor and patch numbers, and breaking changes can only be introduced in versions which change the left-most non-zero component (e.g., going from 0.0.2 to 0.1.0, or from 2.7.3 to 3.0.0 are considered breaking changes).

\item \texttt{Manifest.toml}: this file is automatically generated by the package manager when instantiating an environment, if not already present, and it captures all packages in the environment with their versions, including all indirect dependencies.  When used together with \texttt{Project.toml}, \texttt{Manifest.toml} file describes an exact environment that can be recreated on any other platforms, which is particularly important for reproducibility of applications (e.g., analysis pipelines).
\end{itemize}

Julia packages are organized in directories (usually also Git repositories) in which there is a \texttt{Project.toml} file to define its environment.  Packages can be installed either via path (local path on a machine, or URL to a remote Git repository), or by name if they are present in a package registry. \juliapkg{Pkg.jl} is able to deal with multiple registries at the same time, which can be both public and private, in case there is a need to provide packages relevant only to a single group or collaboration.

Because there is only one package manager and only one way to record package dependency, the Julia package registry simply records the dependency and compatibility metadata separately from package's source code. It allows a local resolver to correctly resolve compatibility in a short amount of time.

Users can interact with the package manager either by using its programmatic API (useful for scripting) or an interactive REPL mode, which can be entered by typing the closing square bracket \texttt{]} in Julia's REPL.  The package manager can also be used to run the tests for packages with a single command (\juliapkg{Pkg.test} using the API, or the command \texttt{test} in the REPL).  Since \juliapkg{Pkg.jl} is a standard library and has many capabilities, all users are familiar with it and do not need to resort to third-party, mutually incompatible tools.

\subsection{Multiple dispatch and Fast Development}

We group multiple dispatch and fast development (i.e., hot reloading) into the same section because of their direct impact on design of the packages and the quality of life of developers.

A side-by-side comparison between OOP and multiple dispatch has been shown in Sec. \ref{polymorphsim}. Here, we highlight that multiple dispatch is also a known solution to the {\em expression problem}~\cite{bib:exprpb,Reynolds1978}. Essentially, in class-based OOP, one of the following is much less natural than the other:
\begin{itemize}
    \item add new methods to existing data type (class/struct).
    \item add new data type (class/struct) for existing algorithm (method).
\end{itemize}

In OOP, the second one is easy, think inheritance; but the first requires access to source code. In Julia, the first one is trivial since methods do not bind to data type (class/struct) to begin with. But the second one can also be easily done by sub-typing the upstream abstract type.

Making it easy for developers to reuse existing packages is crucial in HEP also because libraries are sometimes under-maintained. If we can cleanly extend and reuse these libraries without making private forks, overall efficiency would be boosted.

As a dynamic language, hot reloading should not come as a surprise. It is, in fact, crucial for Julia, due to the inevitable latency introduced by the JIT compiler. The go-to package for this is \juliapkg{Revise.jl}~\cite{bib:revisejl} which automatically detects file modifications and re-compiles the relevant functions on the fly. It can also reload the source code of any Julia Base module, saving a lot of time if (re)compiling Julia.

\subsection{Automatic differentiation} \label{autodiff}

The multiple dispatch system and the native speed of Julia eliminated the need for many specialized libraries to implement the same interface (e.g \codename{NumPy}-interface in \codename{JAX}~\cite{bib:jax}, \codename{TensorFlow}~\cite{bib:tensorflow}, \codename{PyTorch}~\cite{bib:pytorch}). Instead, package maintainers only have to focus on providing rules for the built-in functions as they are fast already and downstream packages are mostly pure-Julia too, see \juliapkg{ChainRules.jl}~\cite{bib:chainrulesjl}. A dedicated organization, JuliaDiff~\cite{bib:juliadiff} collects all the packages and efforts regarding what each autodiff engine is good at.

\section{Foundation HEP-specific libraries to be developed or consolidated}

\subsection{Integration in the ROOT framework}

Because of the ubiquity of \codename{ROOT} in HEP, a Julia interface to this framework, similar to the existing Python one and that will allow people familiar to it to find their way easily, is essential for the development of Julia in the HEP community. In addition, this will provide access to a large set of software used in HEP (storage support, statistics tools, unfolding, etc.) before their counterpart are implemented in Julia. 

\subsection{HEP-specific data storage format}

It will be important to consolidate the support for the \codename{ROOT} data format. The \codename{ROOT} data format is very versatile and allows the storage of instances of arbitrary \Cpp classes (this is true of the current \codename{TTree} and the new \codename{RNTuple} format). Current Julia packages for \codename{ROOT} I/O do not cover this whole versatility and do not allow for reading and writing files with objects of sophisticated types.

\subsection{Physics object types and histogram}

Packages to manipulate Lorentz vectors and to build histograms are already available~\cite{LorentzVectorHEP,FHist}. Leveraging multiple dispatch, these packages are relatively easy to implement, and compose well with rest of the Julia ecosystem (e.g., collection of 4-vectors can be stored and sorted efficiently without any special care). Defining a standard interface to Lorentz vectors and histogram data structures, with a similar approach as the Table interface~\cite{bib:table_jl} could be beneficial. 

\subsection{HEP specific statistical tools}

Over years, HEP community has developed its statistical standard to assert a level of confidence of the experimental results, for measurements, limits and observation of new phenomena. The Julia ecosystem contains several high-quality packages for Bayesian statistics and inference. Two examples are \codename{BAT.jl} (Bayesian Analysis Toolkit in Julia) \codename{BAT.jl}~\cite{bib:batjl} and \codename{Turing.jl}~\cite{bib:turingjl}, which already have been used in several physics analyses. Both packages are being actively developed with good communication of the authors across the development teams. Common interfaces~\cite{bib:densityinterface_jl,bib:inversefunctions_jl,bib:changesofvariables_jl} have been established to increase interoperability.

More development is required for the frequentist CLs approach used at LHC~\cite{ATLAS:2011tau,CMS:2012zhx,ATLAS:2012eoa,Cowan:2010js} and based on profiled likelihood fits. The method is derived from the hybrid method  of the same name developed at LEP~\cite{Junk:1999kv,Read:2002hq} and used later at Tevatron~\cite{Fisher:2006zz,Junk:2006fye}. The \Cpp tools typically used by LHC experiments are the \codename{RooFit} (originally developed for the BaBar experiment~\cite{Verkerke:2006,BaBar:1995bns}) and \codename{RooStats} libraries included in the ROOT framework. For multinodal distributions these libraries are used through the  \codename{HistFactory}~\cite{Cranmer:2012sba} or \codename{HiggsCombine}~\cite{CMS-NOTE-2011-005,CMS:2012zhx,bib:combine_github} tools. The \codename{pyhf} package~\cite{Heinrich:2021gyp} provides a pure-Python implementation of \codename{HistFactory} that offers different computational backends to perform the likelihood maximization and is gaining popularity.  The \codename{HistFactory}, \codename{HiggsCombine}, and \codename{pyhf} are standalone tools, for which inputs are provided in text files (XML or JSON). Thanks to the transparent Julia-Python interface, \codename{pyhf} can also be used in a Julia session or code. For a perfect integration and to exploit the language performance, a Julia implementation is desirable. An effort to implement \codename{pyhf} in Julia has already started~\cite{ling_jerry_2022_7435541} and would need to be consolidated.

Histogram unfolding~\cite{bib:Blobelunfoling} is another statistical tool widely used in HEP experiments. It is used to correct from the effect of the finite resolution of the particle detectors in differential cross section measurement. The \codename{TUnfold}~\cite{Schmitt:2012kp} and \codename{RooUnfold}~\cite{Adye:2011gm} are the most commonly used packages. The \codename{RooFitUnfold}~\cite{Brenner:2019lmf} package provides an extension of \codename{RooUnfold}. New techniques to perform unbinned high-dimensional data unfolding has been recently developed~\cite{Andreassen:2019cjw}. Like for CLs, unfolding comes at the last step of a HEP data analysis, and a Julia implementation would be useful.

\section{Limits of the Julia programming language}

\subsection{Language popularity}\label{sec:pop}

Despite its smaller user base than \Cpp and Python, we have found that it is extremely easy to find information on the web, either from Stack Overflow or from dedicated channels, on Discourse, Slack, and Zulip. The community is very collaborative. An annual conference JuliaCon\footnote{\url{https://juliacon.org/}} is boosting this collaboration. In particular, it encourages exchanges between different fields, both from Academia and Industry. The popularity of Julia is growing and it has been adopted by large academic projects, like the Climate Modeling Alliance (CliMA); and companies like ASML, the largest supplier of photolithography systems; Pharmacology actors like Pfizer, Moderna, and AztraZeneca~\cite{bib:julia-pfizer,bib:husain22,bib:julia-aztrazeneca}; finance actors like Aviva, one of the largest insurers, and the Federal Reserve Bank of New York~\cite{bib:julia-finance, bib:julia-aviva,bib:julia-newyorkfed}.
	
\subsection{Just-in-time compilation latency}\label{sec:latency}

While applications written in Julia run faster than with an interpreted language, the first execution requires additional time to perform the just-in-time (JIT) compilation. In order to limit this overhead, the intermediate results of the compilation, called precompiled code, is cached on disk. The precompilation of a package code is typically performed in parallel at installation time, and the cached content includes, but is not limited to: lowered code, type inference result, etc.; but at the time of writing, Julia does not yet cache compiled machine code\footnote{Progress is being made, see \url{https://github.com/JuliaLang/julia/pull/44527}}. The latency is often called ``time-to-first-plot''.

The JIT compilation latency has been improved from version-to-version, in particular with versions 1.5, 1.6, 1.8, and 1.9 by reducing the number of required recompilations. The various sources of latency have been studied extensively~\cite{bib:Holy2020, bib:Holy2021} and the reduction of the time-to-first-plot is a high priority for the compiler team. Besides improvements coming from the compiler, following the general guidelines of Julia code style for performance~\cite{bib:juliamanual}, which ensure that the compiler can easily infer variable types, should reduce such latency~\cite{bib:Holy2020}. At the same time, tools have been developed to both help ``hunt'' down unnecessary recompilation (\juliapkg{SnoopCompile.jl}~\cite{bib:snoopcompile}), as well as help precompile known common user routinges at installation time (\juliapkg{PrecompileTools.jl}~\cite{bib:precompiletools}).

The latency can also be drastically reduced by preparing a custom system image: the system image contains cached machine code for a set of precompiled packages and past executions. It comes with the drawback that versions of the packages shipped in the system image take precedence over the ones installed via the package manager~\cite{bib:custom_image}, which can be confusing and be a source of bugs. Updating these packages requires rebuilding the custom system image.

The time to produce a first plot, consisting of a 2-D plot of 100 points, was measured to be 2.09$\pm0.01\,$s with Julia 1.9.0-rc1 and the \juliapkg{Plots.jl}~\cite{bib:plots} package. The \juliapkg{Makie.jl} package took 7.57$\pm0.02\,$s using the Cairo backend. Time is similar with the GL backend. Subsequent plots take less than a millisecond. Building a custom system image brings down the latency to below 50$\,$ms for both packages. While for \juliapkg{Plots.jl}, the latency using the standard system image is acceptable, building a custom system image would make use of \juliapkg{Makie.jl} for an interactive session or for a short batch script much more convenient. To measure the improvement brought by the efforts of Julia developers, the measurement is repeated with the older long-term-support release 1.6.7. With this older release the result is 29.0$\pm0.1\,s$ for \codename{Makie.jl}, showing an improvement larger than a factor of 4.

The start-up time could be a concern for large HEP experiment simulation and reconstruction software. As an example of software size, the CMS experiment software, CMSSW~\cite{bib:cmssw}, totals more than 2 million of lines of \Cpp code. The assessment was done with release 12.3.5 and the number of lines of code was defined as the number of semicolons contained in the code. In lieu of a similar sized HEP software package written in Julia, we have measured the start-up time on the relatively large package \codename{OrdinaryDiffEq}, using its version 6.49.4. The package consists of about 125,000 lines of Julia code, excluding comments, and 390,000 when including the external packages. The lines of code have been counted with the \codename{Tokei} software~\cite{bib:tokei} version 12.1.2. and the extra time to run the Example 1 of the manual~\cite{bib:OrdinaryDiffEqDoc} the first time, compared to subsequent executions, was 5.91$\pm0.01\,$s. It goes down to 826$\pm2\,$ms when using a custom system image. We should also note that the precompilation happening on package installation for the package and its dependencies (120 packages) took 256$\,$s only.

For a large experiment software framework, attention will need to be paid to limit code invalidation by respecting the guidelines to ease type inference. This will also help the compiler to optimize the code. While minimizing start-up time may require some effort for large HEP project, we do not expect it to be a show stopper. At worst, it will require to use custom images, with a conciliation on the package management. In addition, development to improve the start-up time is on-going and we should expect significant progress in the near future~\cite{bib:Holy2022}.

\subsection{Application programming interface specification}

Julia lacks a single standard to define the application programming interface (API) of a package. The one with the best support is the use of the \texttt{export} directive to list the symbols exposed to the user. The directive is recognized by the language's introspection functions. The \texttt{names} function lists, by default, the exported symbols, with an option to list all symbols. The \texttt{methodswith} function, used to retrieve functions with an argument of a given type will list only functions from the export list.

Nevertheless, the \texttt{export} directive has the side-effect that all the public symbols end up in the user's namespace, if the package is imported with the \texttt{using} statement. To quote the Julia manual~\cite{bib:juliamanual}, ``it is common to export names which form part of the API. [...] Also, some modules don't export names at all. This is usually done if they use common words, such as derivative, in their API, which could easily clash with the export lists of other modules.''.

The Julia language itself uses the user manual to define the API, as explained in the ``Frequently asked questions'' of this document~\cite{bib:juliamanual}. With such an approach, we lose the benefit of the introspection functions, themselves agnostic to the API information.

A built-in \texttt{unexport} directive, that would allow listing public symbols that the {\tt using} statement must keep in the module namespace and which would be recognized by the introspection functions and also by the documentation generator~\cite{bib:documenter}, would be very beneficial.

\section{Training and language transition support}

Julia has been successfully introduced into existing teams, gradually replacing their \Cpp with Julia packages over time, for example in the LEGEND and BAT groups at the Max-Planck-Institute for Physics. Julia is also the official secondary language (after Python) of the whole LEGEND~\cite{LEGEND:2021bnm} collaboration.

Observed experience is that students with a basic programming background (e.g., in Python or \Cpp) do learn the language very quickly and become productive after just a few days. After exposure to the language for a few weeks, students are typically able to make contributions to larger software packages as well. No problems have been found using Julia for short-term thesis work (e.g., three-month bachelor theses) and even two-week internship, with students and interns who were new to the language. The reaction of these students has been uniformly positive.

Master and PhD theses that used Julia as the primary language have resulted in very positive experience for both students and supervisors. Students who use Julia in longer-term projects not only become very proficient in the language, but also gain a lower-level understanding of computing, data structures and performance implications of modern hardware in general, compared to students who work in Python. This is because Julia makes it very easy the move between higher-level and lower-level programming, in contrast to the Python-plus-\Cpp two-language approach.

More code reuse and transfer has been observed across student generations in Julia, compared to \Cpp. This is due to the combination of an excellent package management with the use of multiple dispatch as a foundation. The first simplifies the maintenance of systems consisting of smaller and more modular packages, while the second solves the {\em expression problem}.

\section{Conclusions}

The Julia programming language has been presented and compared with \Cpp and Python. To study the potential of Julia for HEP, a list of requirements for offline and software-based-trigger HEP applications, covering both the language and its ecosystem, has been established. The compatibility of Julia with these requirements has been studied. Julia and its ecosystem are impressively fulfilling all these requirements. Moreover, Julia brings other features---integrated packaging system with reproducibility support, multiple dispatch and automatic differentiation---from which HEP applications would benefit. 

The capacity to provide, at the same time, ease of programming and performance makes Julia the ideal programming language for HEP data analysis and is more generally an important asset for all the considered HEP applications.  The dynamic multiple dispatch paradigm of Julia has proven to ease code reuse. This property will greatly benefit HEP community applications that involve code developed by many people from many different groups.

Using a single and easy programming language will facilitate training. Experience has shown students with either a \Cpp or Python background learn the language very quickly, being productive after a few days. Using Julia as mainstream language in a collaboration allows students on short-term projects to use the common programming language, while in case of \Cpp, using a simpler language as Python is often needed. This eases the reuse of the code developed in such context.

We have measured the performance provided by the language in the context of HEP data analysis. The measurements show excellent runtime performance, competitive with \Cpp: 11\% faster for the simple LHC event analysis example used as benchmark. When compared to Python, in addition to being faster, it is much less sensitive to implementation choices. The Python implementation was shown to be three orders of magnitude slower than Julia when the event loop is performed in Python. Vectorization techniques can be used to move the event loop by using underlying compiled libraries and this reduces the gap in performance.

One difference with \Cpp and Python is that Julia is younger and has a smaller community. The Julia community is very collaborative and, despite its lower popularity, information for developing with this language is easy to find on the Internet. Julia's rapid growth in academia and industry gives us confidence on the long term continuity of the Julia language, which is essential for HEP projects, because of their large time span.

In view of this study, the HEP community will definitively benefit from a large scale adoption of the Julia programming language for its software development. Consolidation of HEP-specific foundation libraries will be essential to ease this adoption.

\section{Appendix A}
\label{appendix}

\subsection{Polymorphism in \Cpp and Julia illustrated in code}

Differences of polymorphism support in \Cpp, Julia, and \codename{Python} are discussed the ``\nameref{polymorphsim}'' section. This appendix provides code examples illustrating the discussion. 

Static ad-hoc function polymorphism can be implemented in \Cpp using two different paradigms, function overriding and templates. We will illustrate this with an example. Let us consider two classes, \mintinline{c++}{A} and a derived class \mintinline{c++}{AChild}, and a global function \mintinline{c++}{f()}. Using the function overriding paradigm, the ad-hoc polymorphism on the function argument can be implemented as,

\vspace{0.8em}
\begin{minted}[autogobble]{c++}
#include <iostream>

struct A{};
struct AChild: public A{};
struct B{};

void f(A a){
  std::cout << "A\n";
}

void f(B b){
  std::cout << "B\n";
}
\end{minted}
\vspace{0.8em}

The same ad-hoc polymorphism can be implemented in the template paradigm as follows.

\vspace{0.8em}
\begin{minted}[autogobble]{c++}
#include <iostream>
#include <concepts>

template<typename C, typename P>
concept Derived = 
     std::is_base_of<P, C>::value;

struct A{};
struct AChild: public A{};
struct B{};

template<typename T> void f(T x){}

template<Derived<A> T>
void f(T a){
  std::cout << "A\n";
}

template<>
void f(B b){
  std::cout << "B\n";
}
\end{minted}
\vspace{0.8em}

Both implementations can be tested with the following code, which will result in the same output.

\vspace{0.8em}

\begin{tabular}{p{0.7\linewidth}|p{0.3\linewidth}}
\begin{minipage}[t]{\linewidth}
\begin{minted}[autogobble]{c++}
A a;
AChild aChild;
B b;

int main(){
  f(a); //prints A                                                                  
  f(b); //prints B                                                                  
  f(aChild); // prints A                                                           
  return 0;
}
\end{minted}
\end{minipage} & 
                 \begin{minipage}[t]{\linewidth}
                   {\em Output:}
\begin{verbatim}
A
B
A
\end{verbatim}
\end{minipage}
\end{tabular}

\vspace{0.8em}

While in \Cpp, an ad-hoc function polymorphism can be implemented using two different paradigms, the multiple dispatch feature of the Julia language provides a single and consistent way to implement polymorphism. The Julia implementation looks like the following.

\vspace{0.8em}
\begin{minted}[autogobble]{julia}
abstract type AbstractA end
abstract type AbstractAChild <: AbstractA end

struct A <: AbstractA end
struct AChild <: AbstractAChild end
struct B end

function f(a ::AbstractA)
  println("A")
end

function f(b:: B)
  println("B")
end

a = A()
aChild = AChild()
b = B()

f(a) #prints A                                                                  
f(b) #prints B                                                                  
f(aChild) #prints A                                                           
\end{minted}

\vspace{0.8em}

In \Cpp, the dispatch on argument type can be static or dynamic for the class instance argument (\mintinline{c++}{this}) and is always static for the other arguments. This situation can be intricate as in the example below, where the selection of a static or dynamic dispatch over the \mintinline{c++}{this} argument depends on the type of the other argument.

\vspace{0.8em}
\begin{minted}[autogobble]{c++}
#include <iostream>

struct A{
  virtual void f(int x) const { 
    std::cout << "A::f(int)\n"; 
  }
	
  void f(const char* x) const {
    std::cout << "A::f(const char*)\n";
  }
};

struct B: public A{
  void f(int x) const {
    std::cout << "B::f(int)\n";
  }
	
  void f(const char* x) const {
    std::cout << "B::f(const char*)\n";
  }
};

int main(){
  B b;
	
  A& a = b;
	
  a.f(1);  //prints B::f(int)
  a.f(""); //prints A::f(const char*)	
}
\end{minted}
\vspace{0.8em}

The \Cpp language includes an implicit type conversion when copying an object. This feature can lead to confusion, as illustrated in the code below.

\vspace{0.8em}
\begin{minted}[autogobble]{c++}
#include <iostream>

struct Animal{
  virtual void f() const {
    std::cout << "Animal\n";
  }
};

struct Lion: public Animal {
  void f() const {
    std::cout << "King of Savannah\n";
  }
};


void g(const Animal& a){
  a.f();
}

void h(Animal a){
  a.f();
}


int main(){
  Lion leo;
	
  g(leo); //prints King of Savannah
  h(leo); //prints Animal
	
  return 0;
}
\end{minted}
\vspace{0.8em}

This pitfall does not exist in Julia. There is a single way to pass arguments, ``call by sharing''~\cite{bib:wikievalstrategy}, which does not copy the arguments. 

\vspace{0.8em}
\begin{minted}[autogobble]{julia}
		
abstract type Animal end

struct Lion <: Animal 
end

function f(x::Animal)
  println("Animal")
end

function f(x::Lion)
  println("King of Savannah")
end

function g(x)
  f(x)
end

leo =  Lion()
g(leo) # prints "King of Savannah"
\end{minted}
\vspace{0.8em}

The lion is grateful to Julia for honoring his title.

\section{Acknowledgements}

The authors would like to thank Enrico Guiraud (Princeton University and CERN) for the review of the \Cpp code used in the dimuon analysis benchmark and fruitful discussions.

\section{Endorsement}

Johannes Blaschke\footnotemark[1], Ankur Dhar\footnotemark[2], Matthew Feickert\footnotemark[3], Sam Foreman, Cornelius Grunwald\footnotemark[5], Alexander Held\footnotemark[3], Philip Ilten\footnotemark[6], Adam L. Lyon\footnotemark[7], Mark Neubauer\footnotemark[8], Ianna Osborne\footnotemark[9], Johannes Schumann\footnotemark[10], Daniel Spitzbart\footnotemark[11] James Simone\footnotemark[7], Rongkun Wang\footnotemark[12], Michael Wilkinson\footnotemark[6], and Efe Yazgan\footnotemark[13] from the High Energy Physics and Astrophysics communities endorse this work. The authors thank them for their support.

\footnotetext[1]{NERSC, Lawrence Berkeley National Laboratory}
\footnotetext[2]{SLAC National Accelerator Laboratory}
\footnotetext[3]{University of Wisconsin–Madison, Madison}
\footnotetext[4]{Fakultät Physik, Technische Universität Dortmund}
\footnotetext[5]{Massachusetts Institute of Technology,}
\footnotetext[6]{University of Cincinnati}
\footnotetext[7]{Fermi National Accelerator Laboratory}
\footnotetext[8]{University of Illinois Urbana-Champaign}
\footnotetext[9]{Princeton University}
\footnotetext[10]{Erlangen Centre for Astroparticle Physics, Friedrich-Alexander-Universität Erlangen-Nürnberg}
\footnotetext[11]{Boston University, Boston}
\footnotetext[12]{Harvard University, Cambridge}
\footnotetext[13]{National Taiwan University}

\bibliography{julia4hep}

\end{document}